\documentclass[]{aa} 
\usepackage{graphicx}
\usepackage{natbib}
\bibpunct{(}{)}{;}{a}{}{,} 
\newcommand{\TO}[0]{{-}}
\newcommand{\RM}[1]{\mathrm{#1}}
\newcommand{\NOTE}[1]{^{\mathrm{{\it #1}}}}
\newcommand{\kkms}[0]{\RM{K}\,\RM{km}\,\RM{s}^{-1}}
\newcommand{\kms}[0]{\,\RM{km}\,\RM{s}^{-1}}
\newcommand{\micron}{\mbox{$\mu$m}}
\def\thCO{$^{13}$CO}
\def\twCO{$^{12}$CO}

\newcommand{\CIIw}{\ion{C}{ii}}
\newcommand{\CII}{[\ion{C}{ii}]}
\newcommand{\CIw}{\ion{C}{i}}
\newcommand{\CI}{[\ion{C}{i}]}
\newcommand{\OIw}{\ion{O}{i}}
\newcommand{\OI}{[\ion{O}{i}]}
\newcommand{\OIIIw}{\ion{O}{iii}}
\newcommand{\OIII}{[\ion{O}{iii}]}
\newcommand{\NIIIw}{\ion{N}{iii}}
\newcommand{\NIII}{[\ion{N}{iii}]}
\newcommand{\HII}{\ion{H}{ii}}
\newcommand{\HI}{\ion{H}{i}}
\newcommand{\NIIw}{\ion{N}{ii}}
\newcommand{\NII}{[\ion{N}{ii}]}
\newcommand{\msol}{M$_{\odot}$}
\newcommand{\cmcub}{cm$^{-3}$}
\newcommand{\cmsq}{cm$^{-2}$}
\begin{document}

\title{Emission of CO, \CIw, and \CIIw\ in W3\,Main}
  
   \author{C.\,Kramer\inst{1} \and
          H.\,Jakob\inst{1} \and 
          B.\,Mookerjea\inst{1} \and
          N.\,Schneider\inst{2} \and
          M.\,Br\"ull\inst{1} \and
          J.\,Stutzki\inst{1}
          }
   \institute{KOSMA, I. Physikalisches Institut,
              Universit\"at zu K\"oln,
              Z\"ulpicher Stra\ss{}e 77,
              50937 K\"oln, Germany \\
              {\tt lastname@ph1.uni-koeln.de} \and
              Observatoire de Bordeaux,
              Universit\'{e} de Bordeaux 1,
              BP 89, 33270 Floirac, France \\
              {\tt schneider@obs.u-bordeaux1.fr}
             }
   \offprints{C.\,Kramer, \email{kramer@ph1.uni-koeln.de}}
   \date{Received: 20.1.04 / Accepted: 3.6.04 }

   \abstract{We used the KOSMA 3m telescope to map the core
     $7'\times5'$ of the Galactic massive star forming region W3\,Main
     in the two fine structure lines of atomic carbon and four mid-$J$
     transitions of CO and $^{13}$CO. The maps are centered on the
     luminous infrared source IRS\,5 for which we obtained ISO/LWS
     data comprising four high-$J$ CO transitions, \CII, and \OI\ at
     $63$ and $145\,\mu$m. In combination with a KAO map of integrated
     line intensities of \CII\ (Howe et al.  1991), this data set
     allows to study the physical structure of the molecular cloud
     interface regions where the occurence of carbon is believed to
     change from C$^+$ to C$^0$, and to CO.  The molecular gas in
     W3\,Main is warmed by the far ultraviolet (FUV) field created by
     more than a dozen OB stars.  Detailed modelling shows that most
     of the observed line intensity ratios and absolute intensities
     are consistent with a clumpy photon dominated region (PDR) of a
     few hundred unresolved clumps per 0.84\,pc beam, filling between
     3 and 9\% of the volume, with a typical clump radius of 0.025\,pc
     ($2.2''$), and typical mass of 0.44\,\msol.  
%
     The high-excitation lines of CO stem from a $100-200$\,K layer, as
     also the \CI\ lines. The bulk of the gas mass is however at lower
     temperatures.   }

   \maketitle
%

\begin{table*}
\caption[]{List of observational parameters. 
The columns give line frequency, main beam efficieny $B_{\RM{eff}}$,
half power beam width, obsering mode, number of mapped positions, 
velocity channel width $\Delta$v
 (the resolution is a factor of $\sim1.4$ larger)
, mean atmospheric zenith opacity 
$\langle\tau^\mathrm{atm}_0\rangle$, and observing period.
The forward efficiency $F_{\RM{eff}}$ from skydips is constant at 0.9.
}
      
\label{obslist}
\begin{tabular}{llccclrrcr}
\noalign{\smallskip} \hline \hline \noalign{\smallskip}

Species & Transition & Frequency & $B_\RM{eff}\NOTE{c}$ & HPBW$\NOTE{c}$ &
  Mode & Pos. & $\Delta$v & $\langle\tau^\RM{atm}_0\rangle$ &
  Observing period \\
  &&[GHz]&&[$''$]&&&[m s$^{-1}$] & & \\
\noalign{\smallskip} \hline \noalign{\smallskip}

$^{12}$CO  & $J=3\TO2$           & 345.7960 & 0.65 & 76 & OTF$\NOTE{a}$ 
    & 1091 & 29 & 0.54 & 11/2000\\
$^{13}$CO  & $J=3\TO2$           & 330.5880 & 0.65 & 76 & OTF 
    & 1091 & 29 & 0.63 & 1/2001, 8/2002 \\
$^{12}$CO  & $J=4\TO3$          & 461.0408 & 0.57 & 57 & DBS$\NOTE{b}$
    &  192 & 675 & 1.09 & 10/2001 \\
$^{12}$CO  & $J=7\TO6$           & 806.6518 & 0.25 & 42 & DBS 
    &  192 & 389 & 1.15 & 12/2001 \\
$^{13}$CO  & $J=8\TO7$           & 881.2728 & 0.38$\NOTE{d}$ 
    & 40$\NOTE{d}$  & DBS 
    &   60 & 356 & 0.89 &  1/2003 \\
$^{12}$\CI & $^3$P$_1\TO^3$P$_0$ & 492.1607 & 0.50 & 55 & DBS 
    &  192 & 634 & 0.82 & 12/2001 \\
$^{12}$\CI & $^3$P$_2\TO^3$P$_1$ & 809.3420 & 0.25 & 42 & DBS 
    &  192 & 388 & 1.10 & 12/2001 \\
\noalign{\smallskip} \hline \end{tabular}
\begin{list}{}{}
\item[$\NOTE{a}$]OTF -- On-The-Fly observing mode using an 
   emission-free off position at $(42.5', -25.7')$ 
\item[$\NOTE{b}$]DBS -- Dual-Beam-Switch mode using the secondary mirror 
   ($6'$ throw in azimuth)
\item[$\NOTE{c}$]The main beam efficiency and the HPBW were
    determined from cross scans of Jupiter.
  \item[$\NOTE{d}$]Data at 880\,GHz were taken after an adjustment of
    the primary mirror following photogrammetric measurements by
    \citet{miller2002}. The beam efficiency at 880\,GHz was
    extrapolated from new efficiencies taken at 810, 492, and 345\,GHz
    using the Ruze formula, consistent with a surface accuracy of
    25\,$\mu$m rms.  The HPBW was scaled using the width at 810\,GHz.
\end{list}
\end{table*}

\section{Introduction}
   
Photon Dominated Regions (PDRs) are surfaces of molecular clouds where
heating and chemistry are dominated by FUV $(6\,{\rm
  eV}<h\nu<13.4\,{\rm eV})$ radiation.  The photons can scatter deep
into the molecular clouds due to their clumpy structure. A large
fraction of the volume and mass is thus photon dominated
\citep{tielens1985}.  While FIR continuum emission by dust is by far
the major coolant, the gas cools mainly via the atomic fine structure
lines of \OIw, \CIIw, \CIw, and the rotational CO lines
\citep[e.g.][]{kaufman1999}.  Carbon bearing species are thus of great
importance. The occurrence of carbon with depth into the surfaces
changes from C$^{+}$ to C$^{0}$ and -finally- to CO. \CII, \CI\ and
mid-$J$ CO lines trace the outer layers of photon dominated molecular
clumps. However, \CII\ emission also stems from \HII\ regions, \HI\ 
clouds (i.e. the cold neutral medium (CNM)
\citep{kulkarni1987,wolfire1995}), and diffuse \HII\ regions (i.e. the
warm ionized medium (WIM) \citep{heiles1994}).  In the context of
PDRs, depending on the induced chemistry, the critical densities, and
energy levels of their transitions, \OI, \CII, \CI, and CO lines serve
as probes to trace different temperature and density regimes.  In
particular the two fine structure lines of neutral atomic carbon at
492 and 809 GHz provide information on the intermediate regions
between atomic and molecular gas while tracing two different physical
regimes.

Few studies have so far tried to map the major cooling lines of
\CIIw/\CIw/CO in individual regions to derive a coherent picture of the
chemical, physical, and -using velocity resolved line profiles-
dynamical structure of the emitting regions.  The studies of the
massive star forming regions S106 by
\citet{schneider2003,schneider2002} and Orion Molecular Cloud 1 by
\citet[][]{mookerjea2003} and \citet[][]{yamamoto2001}, as well as the
study of the reflection nebula NGC\,7023 by \citet{gerin1998} show
that a coherent picture is not easily found using standard,
steady-state PDR models.
Especially the large spatial extension of neutral carbon and the high
abundance deep inside molecular clouds appears to be in contradiction
with standard PDR models.  This is also addressed by the \CI\ vs. CO
studies of \citet{plume1994} in S\,140, \citet{oka2001} in DR15, 
\citet{kamegai2003} in $\rho$ Ophiuchi, and \citet{oka2004} in NGC\,1333.
In a study of the translucent cloud MCLD\,$123.5+24.9$,
\citet{bensch2003} find lower than expected \CI\ intensities and \CI\ 
to CO ratios.  \citet{kamegai2003}
and \citet{oka2004}
argue that \CI\ traces young evolutionary stages of cloud surfaces
which is consistent with time-dependent model calculations
\citep[cf. also][]{stoerzer1997}.  \citet{gerin1998} suggest that a
second PDR deeper in the cloud explains the observed enhanced \CI/[CO]
ratios in NGC7023. The importance of the interclump medium, providing
effective shielding of the CO in the embedded clumps, is discussed by
\citet{bensch2003}.

%
%

To improve our understanding of PDRs in different sources with
different geometries and different FUV radiation fields, we present
here a study of W3\,Main. As the major cooling lines with their bright
transitions are particular important as tracers in external galaxies,
proper and detailed understanding of their emission and the chemical
and physical conditions of the originating gas is necessary as a firm
basis for their interpretation.

The chain of giant \HII\ regions W3-W4-W5 lie at the edge of the Cas
OB6 association and form part of the Perseus spiral arm.
\citet{heyer1998} combined large-scale CO 1--0 FCRAO maps with \HI\ 
line emission and 21\,cm continuum DRAO data \citep{normandeau1997}
\citep[cf.][]{read1981}, and IRAS HIRES images at simmilar angular
resolutions of about $1'$.  W3 is associated with the W3 giant
molecular cloud comprising more than $10^5$\,M$_{\odot}$\ at a
distance of 2.3\,kpc \citep{georgelin_georgelin1976}.  This distance
will be used throughout this paper, though \citet{imai2000} recently
found indications for a distance of only 1.8\,kpc.  W3\,Main is a well
known site of high-mass star formation comprising several \HII\ 
regions
\citep{tieftrunk1995,tieftrunk1997,tieftrunk1998a,tieftrunk1998b}.
ISO/SWS and LWS spectroscopy of the compact \HII\ region W3\,A
\citep{peeters2002} reveals strong atomic emission lines.  W3\,A is
associated with a luminous infrared source IRS\,2, belonging to a
cluster of infrared sources (IRS\,1--7) found by
\citet{wynn1972,richardson1989}. The total luminosity of IRS\,5 is
estimated to be $2-5\,10^5$\,L$_{\odot}$\ 
\citep{werner1980,thronson1980,campbell1995}. \citet{ladd1993}
observed the associated emission of dust in the submillimeter
identifying three cores (SMS\,1$-$3). NIR observations by
\citet{megeath1996} reveal that high-mass star formation is associated
with the formation of a dense cluster of several dozens of low-mass
stars.
%

%
%
%
 Column densities of molecular gas were mapped by \citet{tieftrunk1995}
 in millimetric C$^{18}$O and C$^{34}$S lines and by
 \citet{roberts1997} in $^{13}$CO and C$^{18}$O $1-0$. The latter
 observation were conducted with BIMA and reveal six clumps with
 diameters between 0.1 and 0.3\,pc (masses between 120 and 480
 M$_\odot$) in the vicinity of IRS\,5 and IRS\,4. 
 The denser and warmer molecular gas in the vicinity of
 IRS\,5$-$IRS\,4 was observed by \citet{hasegawa1994} in CO $3-2$ and
 $6-5$. 
%

As part of a larger study of dense, hot gas in Galactic star forming
regions, \citet{kruegel1989} and \citet{boreiko1991} observed
$^{12}$CO $7-6$ and $^{12}$CO, $^{13}$CO $9-8$ towards W3\,IRS\,5,
respectively. The large-scale extent of the 492\,GHz line of atomic
carbon and of $^{13}$CO and C$^{18}$O $2-1$ was observed by
\citet{plume1999} at $3'$ resolution. 
%
%
\citet{howe1991} mapped the \CII\ 158$\,\mu$m emission of W3\,Main
and developed models to reconcile the distribution of \CII\ emission
and the FUV field.
%


Here, we attempt to model the mapped emission of mid-$J$ and high-$J$
CO lines, of both \CI\ lines, and of \CII, at resolutions of $\sim1'$.
Section 2 describes the KOSMA observations and the ISO/LWS data set.
Section 3 presents the maps of integrated intensities, spectra at four
selected positions, and a first derivation of physical properties. In
Sect. 4 we discuss the physical structure of the photon dominated
regions.  We present a model of the FUV field, needed for a detailed
comparison of line integrated intensities with PDR models.  We
summarize our results in Sect. 5.

\begin{figure*}[p]
\centering
\includegraphics[angle=-90,totalheight=7.55cm]{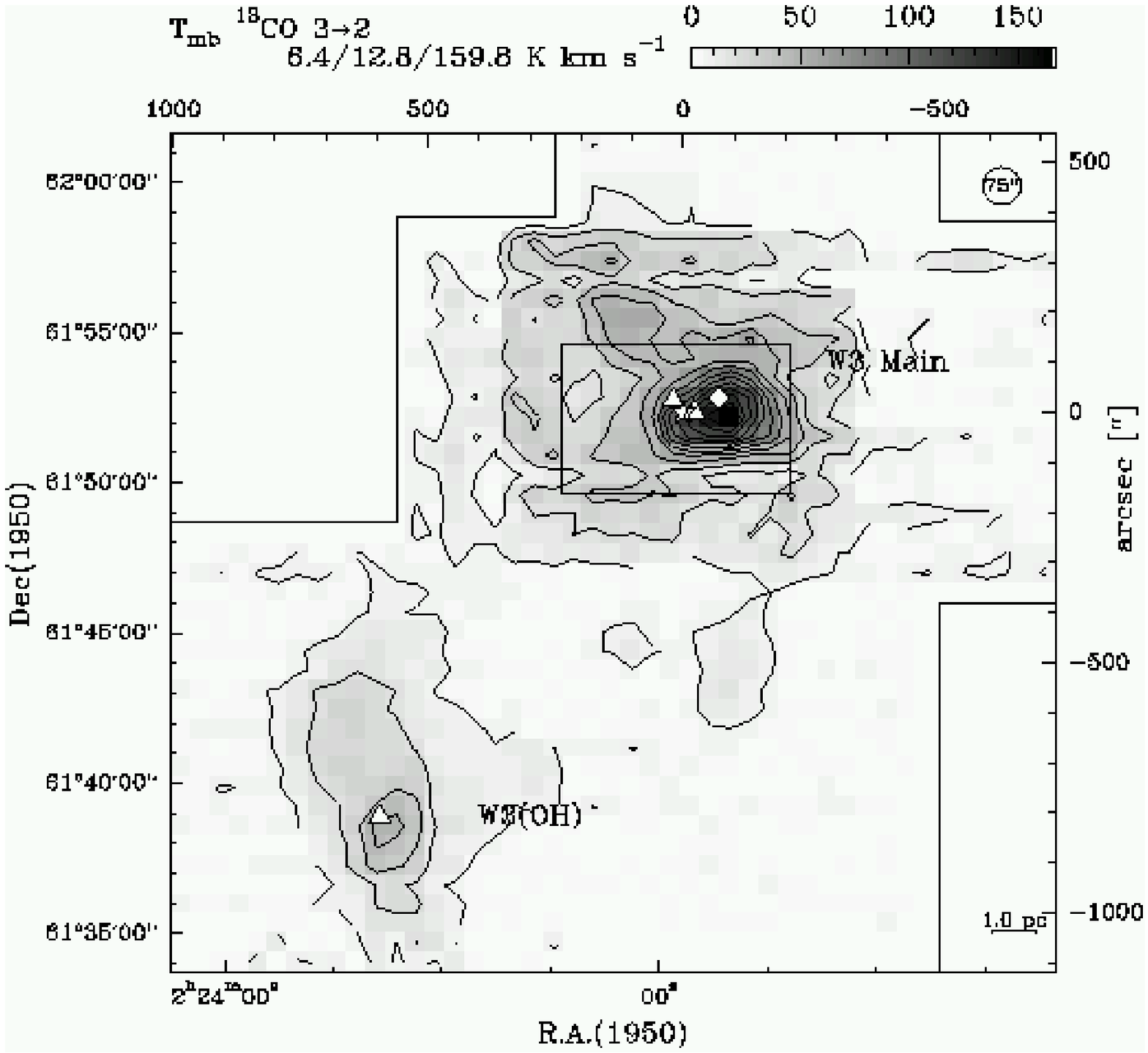}\nobreak
\includegraphics[angle=-90,totalheight=7.55cm]{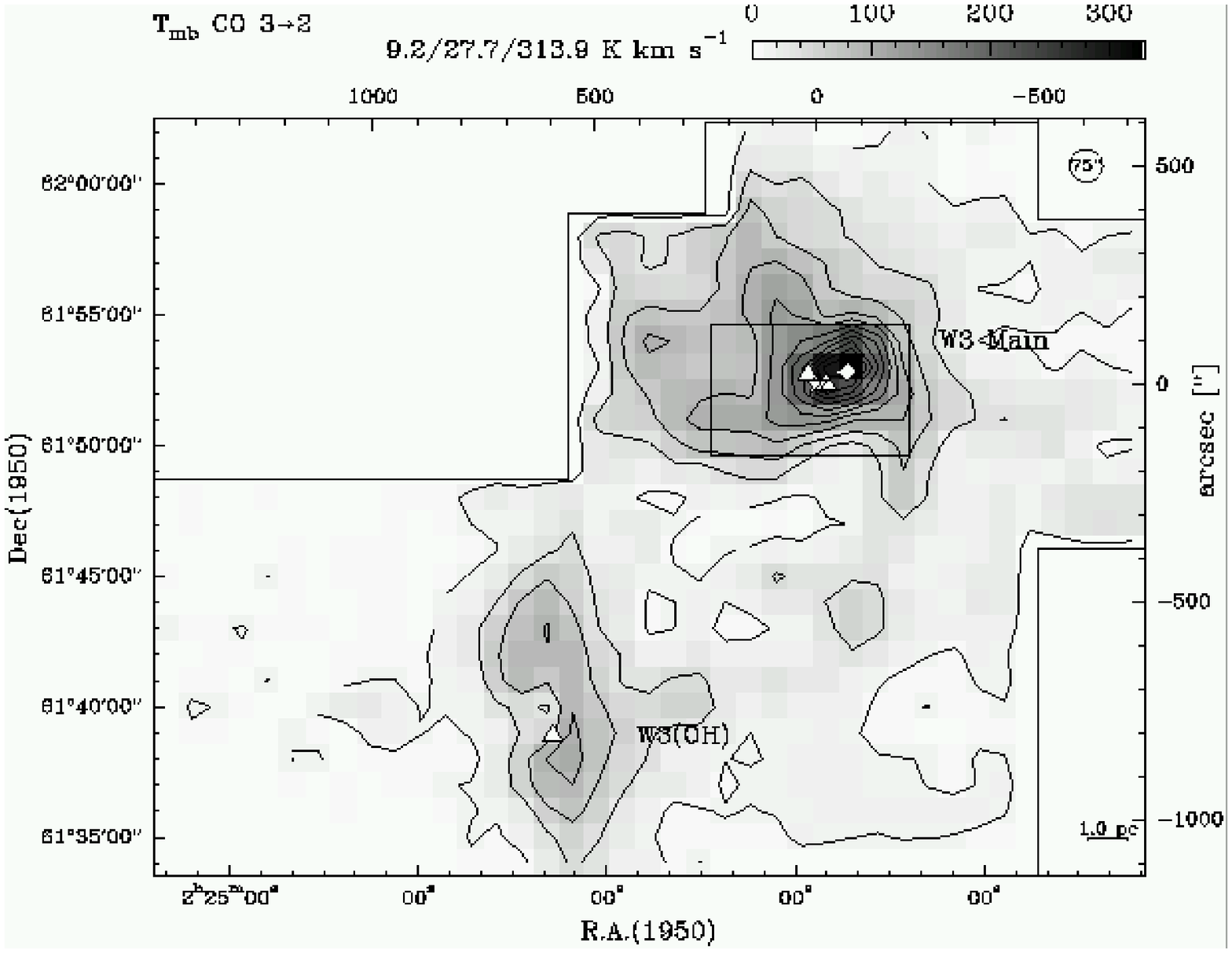}
\caption{{\em (Left)} Integrated $^{13}$CO and {\em (Right)} $^{12}$CO $3\TO2$
  large scale maps of W3, including W3\,Main in the north and W3(OH)
  in the south. The integration interval is [$-55,-25\,\kms$] and thus
  also covers the emission of W3(OH). Offsets are given in arcsec. 
  Contour levels are described in the upper left corner of each map: 
  first contour$=3\,\sigma$/step size/peak intensity.
  The position $(0,0)$ of IRS~5 is marked by a star, IRS~4 is marked
  by a square. Triangles mark the positions of IRS~1 (to the east), IRS~3, and
  W3(OH). The inner black box indicates the size of the
  higher frequency maps shown below.  The 0/0 position is at IRS\,5:
  $\alpha{(1950)}=02\fh21\fm53\fs2$ $\delta{(1950)}=61\degr52\arcmin21\arcsec$.
\label{CO32}
}
\end{figure*}
\begin{figure*}[p]
\centering
\includegraphics[angle=-90,totalheight=6.0cm]{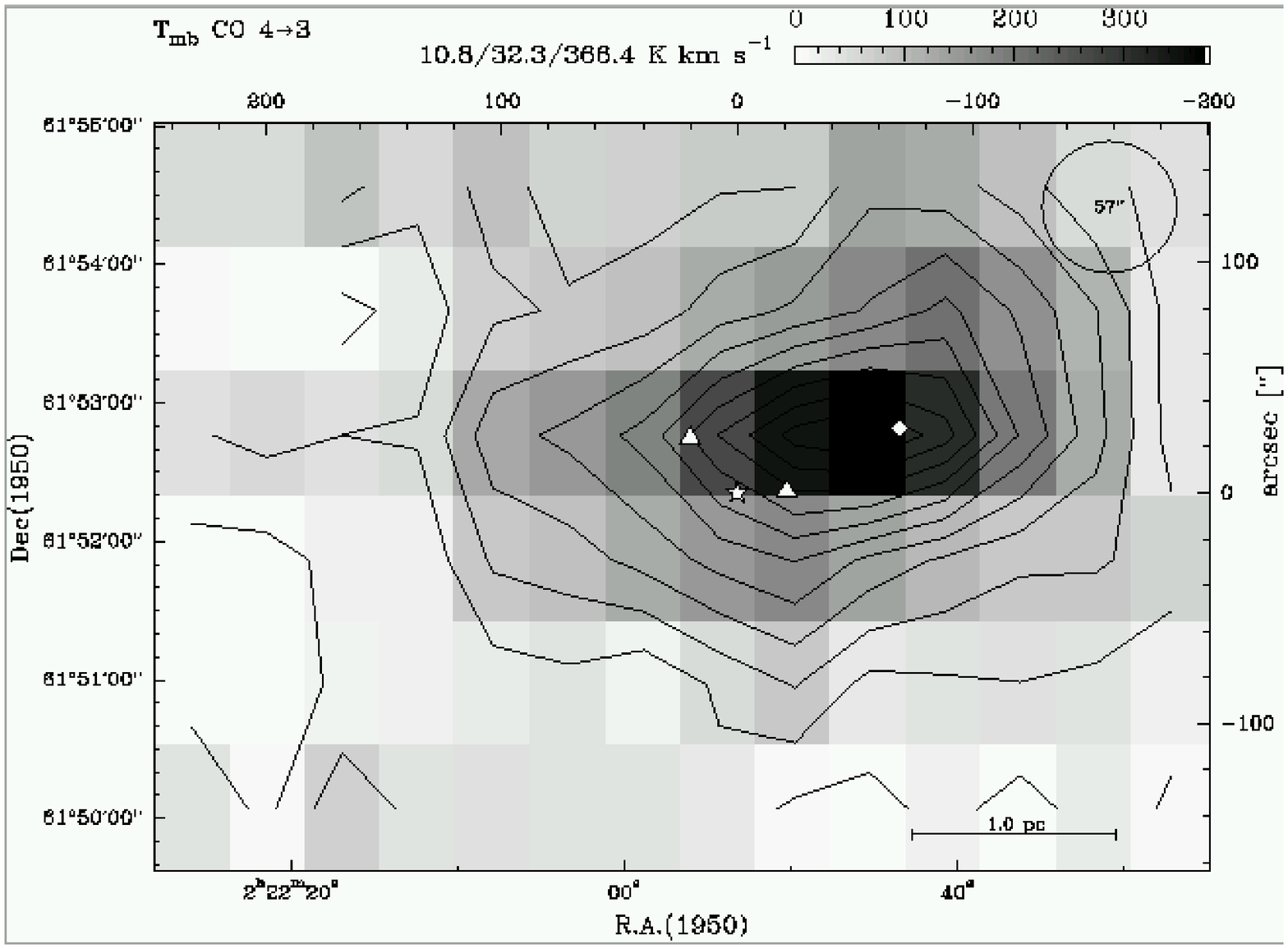}\nobreak
\includegraphics[angle=-90,totalheight=6.0cm]{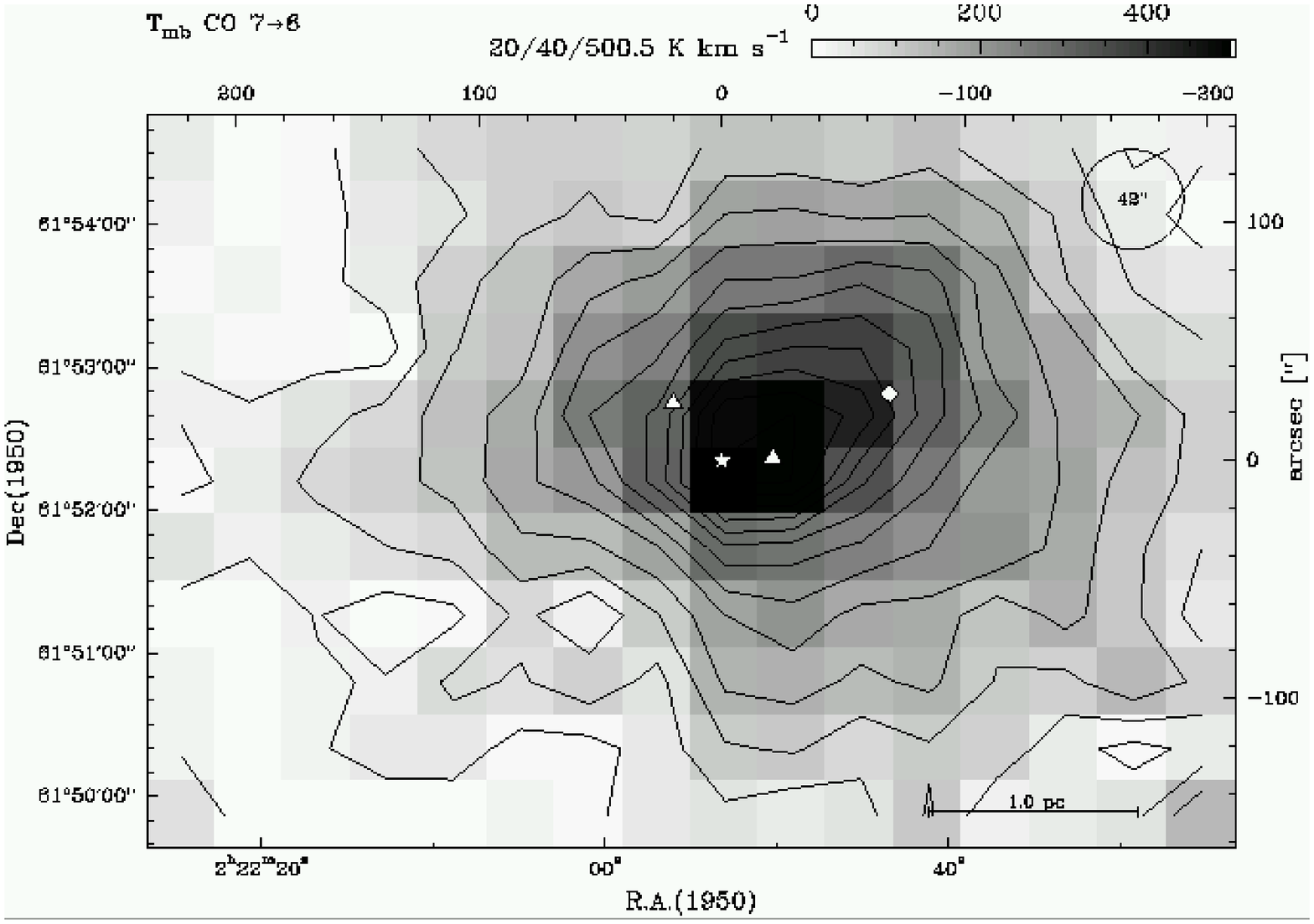}\break
\includegraphics[angle=-90,totalheight=6.0cm]{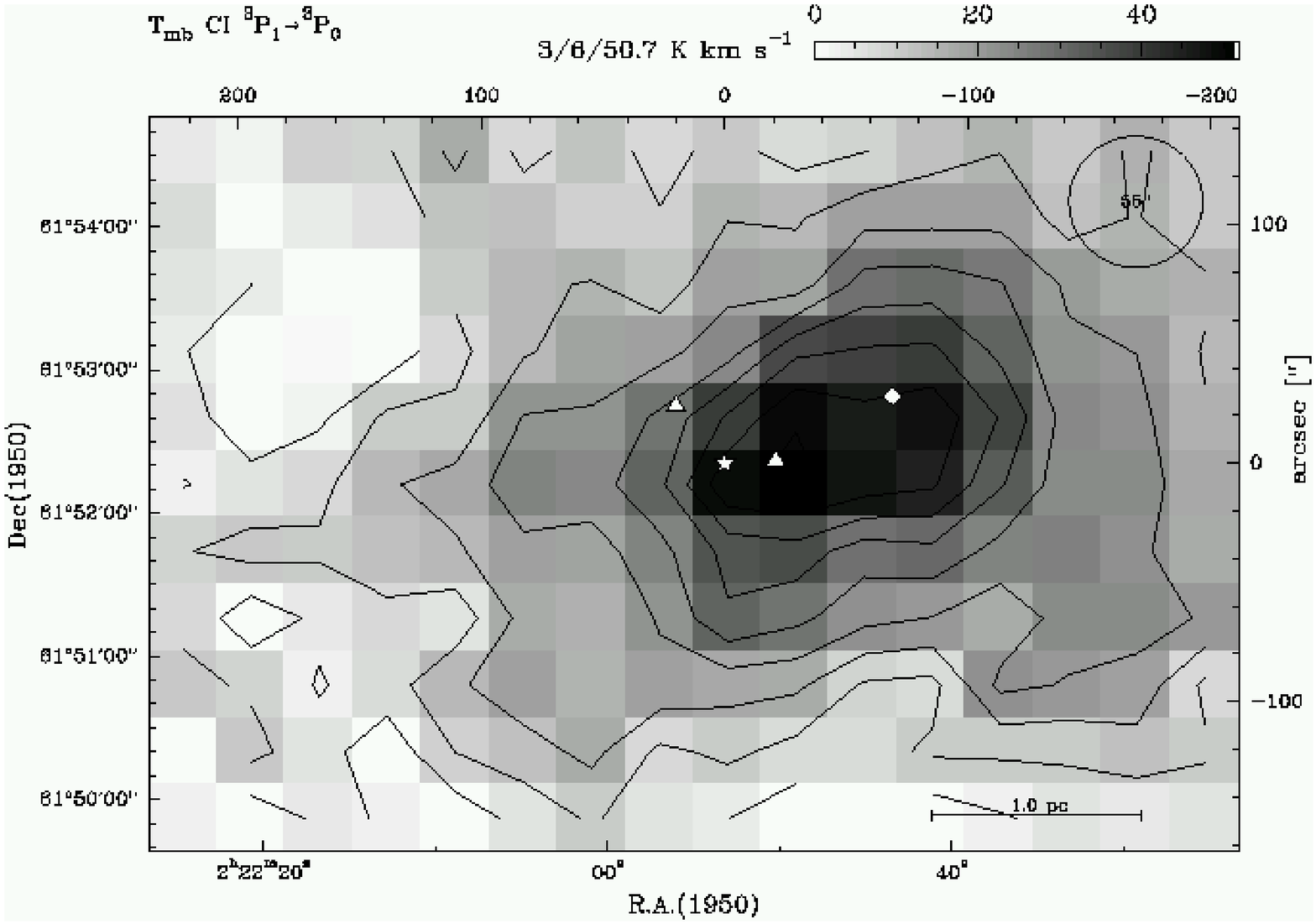}\nobreak
\includegraphics[angle=-90,totalheight=6.0cm]{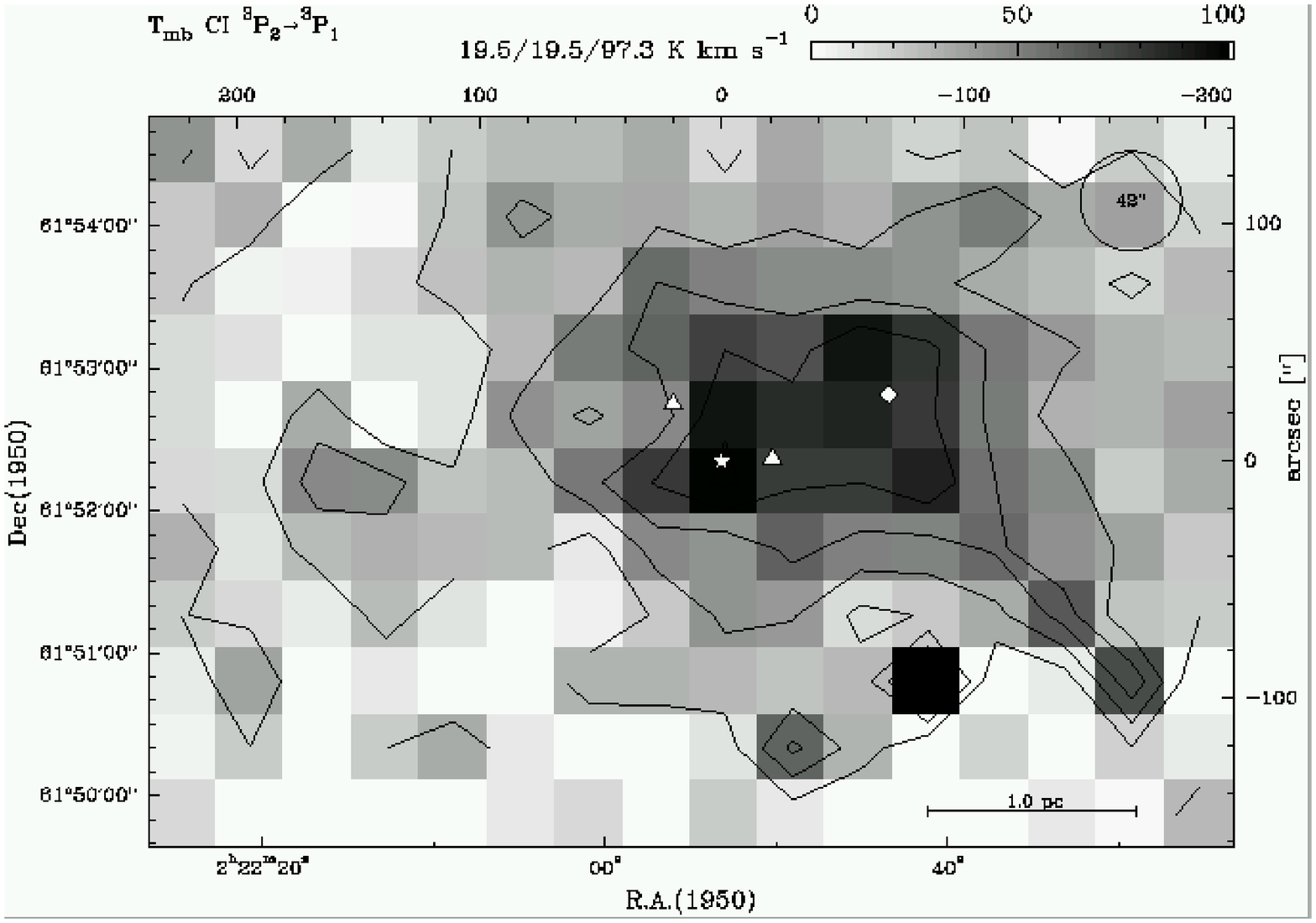}
\caption{{\em (From upper left to lower right)} Integrated intensity
maps of  CO $4\TO3$, CO $7\TO6$, \CI\ $1\TO0$ and \CI\ $2\TO1$
of the W3 Main region (indicated by the rectangle in Figure\,\ref{CO32}).
\label{midj}
}
\end{figure*}


\begin{figure}[htb]
\centering
\includegraphics[angle=-90,width=\linewidth]{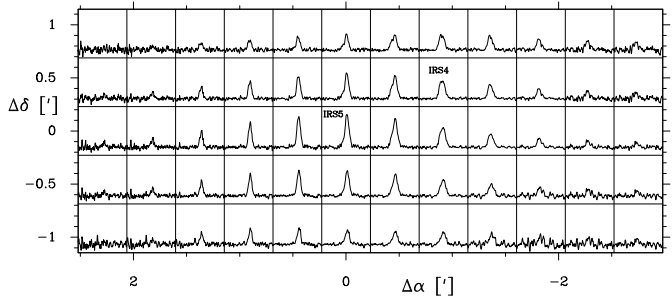}
\caption{Map of $^{13}$CO 8--7 spectra at a resolution of $75''$ (0.84\,pc) and on the
  $T_{\rm mb}$ scale after a pointing correction. The temperature
  range is $-2$ to 10\,K. The $v_{\rm lsr}$ range is $-70$ to
  $-10$\,kms$^{-1}$.
\label{fig-13co87}
}
\end{figure}

\section{Observations} 

The KOSMA 3m submillimeter telescope on Gornergrat, Switzerland
\citep{winnewisser1986,kramer1998kona,kramer2000tucson} was used to
observe the transitions of carbon and the mid-$J$ transitions of CO.

\subsection{[\ion{C}{i}] and mid-$J$ CO}

We have mapped W3 (Fig.\,\ref{CO32}, \ref{midj}) in the two atomic
carbon fine structure transitions (\CI) at 492\,GHz (609~$\mu$m,
$^3$P$_1\TO^3$P$_0$; hereafter $1\TO0$) and 809\,GHz (370~$\mu$m,
$^3$P$_2\TO^3$P$_1$; hereafter $2\TO1$) and the mid-$J$ rotational
transitions $J=4\TO3$ and $J=7\TO6$ of CO, and $J=8\TO7$ of $^{13}$CO
using the newly installed SubMillimeter Array Receiver for Two
frequencies (SMART) \citep{graf2002} on KOSMA.

SMART is a dual-frequency eight-pixel SIS-heterodyne receiver that
simultaneously observes in the 650 and 350\,\micron\ atmospheric
windows \citep{graf2002}. It consists of two 4 pixel subarrays, one
operating in the frequency band of $455-495$ GHz and the other in the
band of $795-882$ GHz. The IF signals are analyzed with two 4-channel
array-acousto-optical spectrometers with a spectral resolution of
1.5\,MHz \citep{horn1999}. The IF frequency of 1.5\,GHz is ideally
suited for simultaneous detections of the [\ion{C}{i}] $2\TO1$ line at
810 GHz and the CO $7\TO6$ line at 806\,GHz in opposite sidebands. The
two subarrays are pointed at the same positions on the sky. Thus, four
spatial positions are observed simultaneously. The [\ion{C}{i}]
$1\TO0$ \& $2\TO1$ and the CO $7\TO6$ observations of W3 presented
here were done simultaneously.  The relative pointing is thus
constant. Due to the simultaneous observations at both frequency
bands, relative calibration errors are strongly reduced and line
ratios are measured precisely \citep[cf. e.g.][]{stutzki1997}.  The typical
receiver noise temperatures achieved at the center of the bandpass is
around 150\,K at 492\,GHz and between 500 and 600\,K at 810\,GHz. The
receiver setup consists of a K-mirror type image rotator, two
Martin-Puplett diplexers, and two solid state local oscillators, which
are multiplexed using collimating Fourier gratings \citep{graf2002}.
The image rotator corrects for the sky rotation during the observation
and it allows to set the array to any desired orientation.

The observations were performed in Dual-Beam-Switch mode (DBS), with a
chop throw of $6'$ in azimuth. This mode of observation produces flat
baselines. Only a zeroth order baseline was subtracted from the
calibrated spectra.  
The spacing between two adjacent pixels is $114''$. Pointing was
frequently checked 
on Jupiter and the sun. The resulting accuracy is better than $\pm20''$.

An area of $\sim7'\times5'$
was mapped in the two [\ion{C}{i}] lines and the two CO lines, on a
$27.5\arcsec\times 27.5\arcsec$ grid with a total integration time of
160\,s per position. Due to the extent of the W3 cloud complex weak
self-chopping artefacts were found at the borders of the mapped
region. This was confirmed by separate observations of these positions
in total power position switch mode. We did not correct for these
artefacts.

Atmospheric calibration was done by measuring the atmospheric emission
at the OFF-position to derive the opacity \citep{hiyama1998}. Spectra
of the two frequency bands near 492 and 810\,GHz were calibrated
separately. Sideband imbalances were corrected using standard
atmospheric models \citep{cernicharo1985}.
  The CO $7\TO6$ line observed in the image sideband, was also
  corrected.


  
  A small map of the $^{13}$CO 8--7 line at 880\,GHz covers an area of
  $5.5'\times2.3'$ centered on IRS\,5 and IRS\,4
  (Fig.\,\ref{fig-13co87}). Integration times are on average 480\,sec
  per position. No self-chopping effects are discernable in the
  resulting spectra. A pointing offset was corrected off-line by
  shifting the $^{13}$CO data by ($0''$,$-27.5''$).
   The main beam efficiency $B_\RM{eff}$ and HPBWs were determined
    using continuum cross scans on Jupiter (Table\,\ref{obslist}).
    \citet[][Table\,1]{griffin1986} observed Jupiter and Mars at 10
    frequencies in the $0.35$ to 3.3\,mm wavelength range using Mars
    as primary standard. We used their Jovian brightness temperatures,
    linearily interpolated, to derive the temperatures at the
    frequencies of the KOSMA observations.  The KOSMA half power
    beamwidths (HPBWs) were derived from the observed FWHMs.  
At 810\,GHz $\sim25$\% of the emission is detected by the main beam.
Continuum scans of the edge and disc of the sun reveal an extended
Gaussian error beam component of $\sim200''$ FWHM, detecting about
$40\%$ of the emission.  We attribute this error beam to residual
misalignment of the $18$ panels of the primary mirror.
  Due to the large extent of the errorbeam relative to the mapped
  region, we cannot convolve out the errorbeam. Instead, we simply
  scaled the antenna temperature data with the ratio of forward
  over main beam efficiency to go from the $T_A^*$ scale to the
  $T_\RM{mb}$ scale.

From observations of reference sources, we estimate the relative
calibration error to be $\sim15\%$.




\subsection{Low-J CO \& \thCO}

We have used a dual-channel SIS receiver \citep{graf1998} on KOSMA to
observe large scale maps in CO $J=3\TO2$ and $^{13}$CO $3\TO2$
emission of W3 \citep{mueller2001}. The observations were done using
the On-The-Fly (OTF) technique \citep{beuther2000,kramer2000tucson} on
a fully sampled grid of 
 $30'\times 30'$.
  For the low-frequency
observations we used the high resolution spectrometer (HRS1) till
Winter 2001 and the variable resolution spectrometer (VRS1) lateron.
Sinusoidal baselines were subtracted.
See Table\,\ref{obslist} for a summary of observational parameters.

\subsection{ISO/LWS spectroscopic data}\label{chapter-obs-iso}
\label{sec-iso-observations}

ISO/LWS \citep{clegg1996} spectra are available from the data archive
for the position of IRS\,5 in W3\,Main (TDT\,47301305). The grating
spectrometer covers the wavelength range 
%
$47-197\,\mu$m.
The standard OLP-10 data product was used along with ISAP v.\,2.1 to
deglitch, defringe, and average the data.  
Linear baselines were subtracted from all spectra.
Gaussian fits, with the line width held fixed at the instrumental
width, were done to derive integrated intensities.

\section{Results} 
\subsection{Why to use different tracers?}

The observed transitions trace varying physical conditions of the
emitting regions. Transitions are characterized by their critical
density for collisions with H$_2$ or other partners and by their upper
level energy \citep[see Table\,2 in][]{kaufman1999}. For the CO
$J\rightarrow{}J-1$ transition, the critical density is $4\,10^3\,J^3$
cm$^{-3}$ and the upper level energy corresponds to $2.8\,J(J+1)$
Kelvin. The mid-$J$ CO transitions observed with KOSMA are thus
sensitive to gas at about $10^5$ to $10^6$ \,cm$^{-3}$ at temperatures
between 30 and 200\,K while the high-$J$ lines traced by ISO/LWS are
sensitive to typically $10^7$\,\cmcub\ and 650\,K. The regime of high
densities $>10^5$\,\cmcub\ and high temperatures $>200\,$K is also
traced by the two atomic \OI\ lines at $145\,\mu$m and $63\,\mu$m. On
the other hand, the two atomic \CI\ lines and the \CII\ line trace
energies of less than 100\,K and densities of less than
$3\,10^3$\,\cmcub.


\subsection{Maps of integrated intensities}

We present here KOSMA maps of W3\,Main at spatial resolutions of about
one arcminute.  The $^{12}$CO and $^{13}$CO $3-2$ maps include
W3\,Main and W3\,(OH), and cover an area of $\sim20'\times14'$.  A
$7'\times5'$ region centered in W3\,Main-IRS5 was mapped in \CI\
$^3$P$_{1}-^3$P$_{0}$, $^3$P$_{2}-^3$P$_{1}$, $^{12}$CO $4-3$, and
$7-6$. The $5.5'\times2.3'$ map of $^{13}$CO $8-7$ covers the IRS\,5
and IRS\,4 positions.
 We have included in our analysis the \CII\ map of \citet{howe1991}
 and ISO/LWS data at the position of IRS\,5.

In contrast to the velocity integrated CO maps which show only one
maximum near IRS\,5, the corresponding spectra show many details like
self-absorption features and overlapping velocity components. These
will be discussed in Sect.\,\ref{chap-spectra}.

\subsubsection{Maps of $^{12}$CO and $^{13}$CO $3-2$ }

The \twCO\ and \thCO\ $3\TO2$ maps exhibit ($i$) a single central peak
close to the position of IRS\,5, ($ii$) an extension to the north-east
toward W3\,North \citep[cf.][]{plume1999}, which is the northernmost
member of the string of active star-forming regions, and ($iii$) a
ridge of emission extending towards the south in the direction of
W3\,(OH). In addition, ($iv$) both maps show another ridge of emission
extending eastward from IRS\,5 and then bending north-east. The two
north-eastern ridges are more clearly seen in the \thCO\ isotopomer
than in \twCO.

\subsubsection{Maps of CO $4-3$ and $7-6$}

The higher-frequency maps cover the central part of W3\,Main
encompassing an area of $\sim7'\times5'$.

The CO $4-3$ and $7-6$ maps again show a strong peak near IRS\,5,
slightly shifted towards IRS\,4, which drops rather steadily in all
directions. A ridge of emission extending from IRS\,4 in south-western
direction is discernable in the CO $7-6$ data set. 



\subsubsection{Maps of [\ion{C}{i}] $1-0$ and $2-1$}



Comparison of the \CIw\ maps with the CO maps in Figures~\ref{CO32}
and \ref{midj} shows that although there is a large-scale similarity
between the two emission patterns, the mid-$J$ CO emission is more
centrally concentrated than the [\ion{C}{i}] emission. This indicates
the origin of the CO emission in regions with high density and
temperature, while the [\ion{C}{i}] emission originates in regions of
comparatively lower density and temperature, and hence traces out the
widespread diffuse UV illuminated neutral gas.  


\subsection{Spectra at selected positions}
\label{chap-spectra}

\begin{figure*}[th]
\centering
\includegraphics[width=4cm]{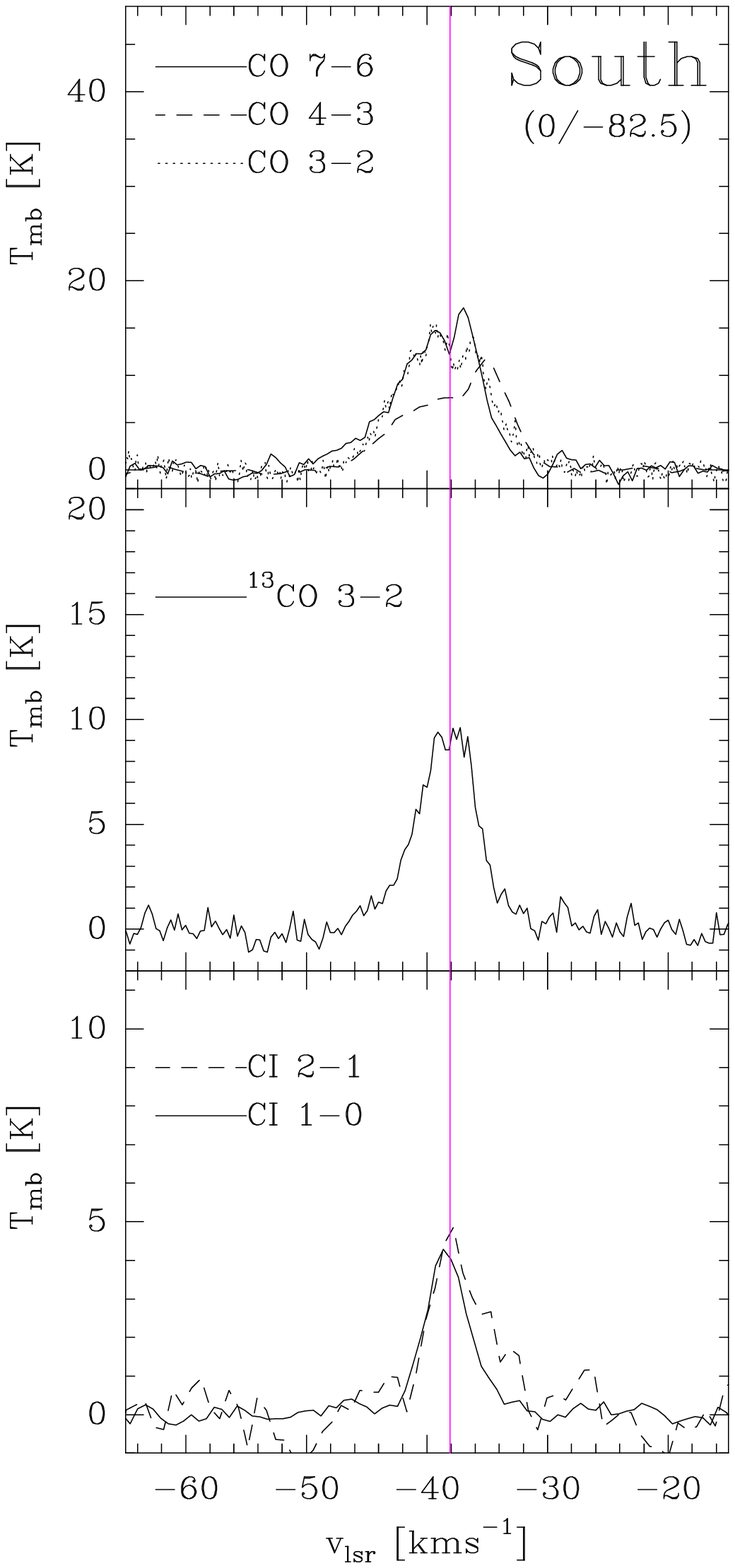}
\includegraphics[width=4cm]{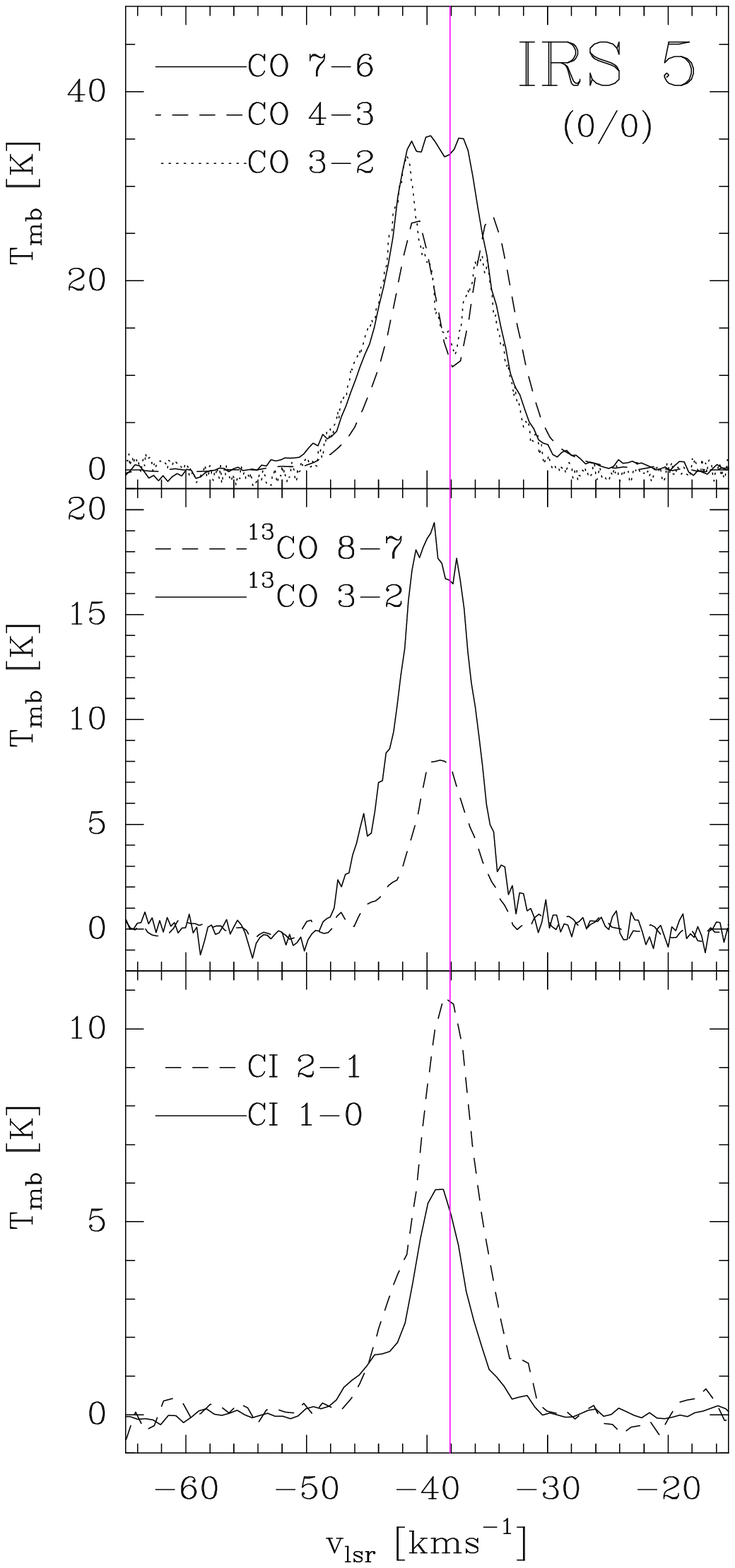}
\includegraphics[width=4cm]{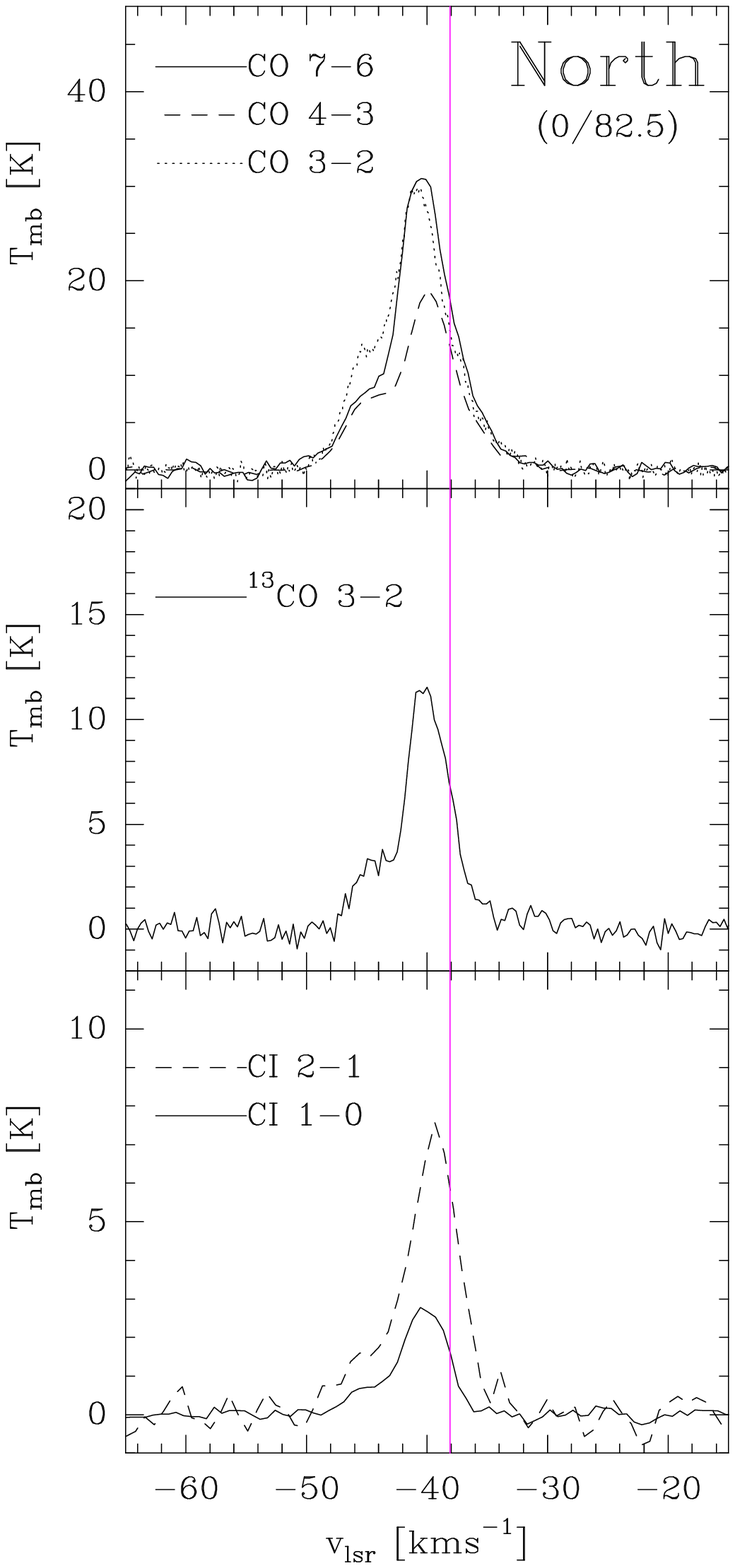}
\includegraphics[width=4cm]{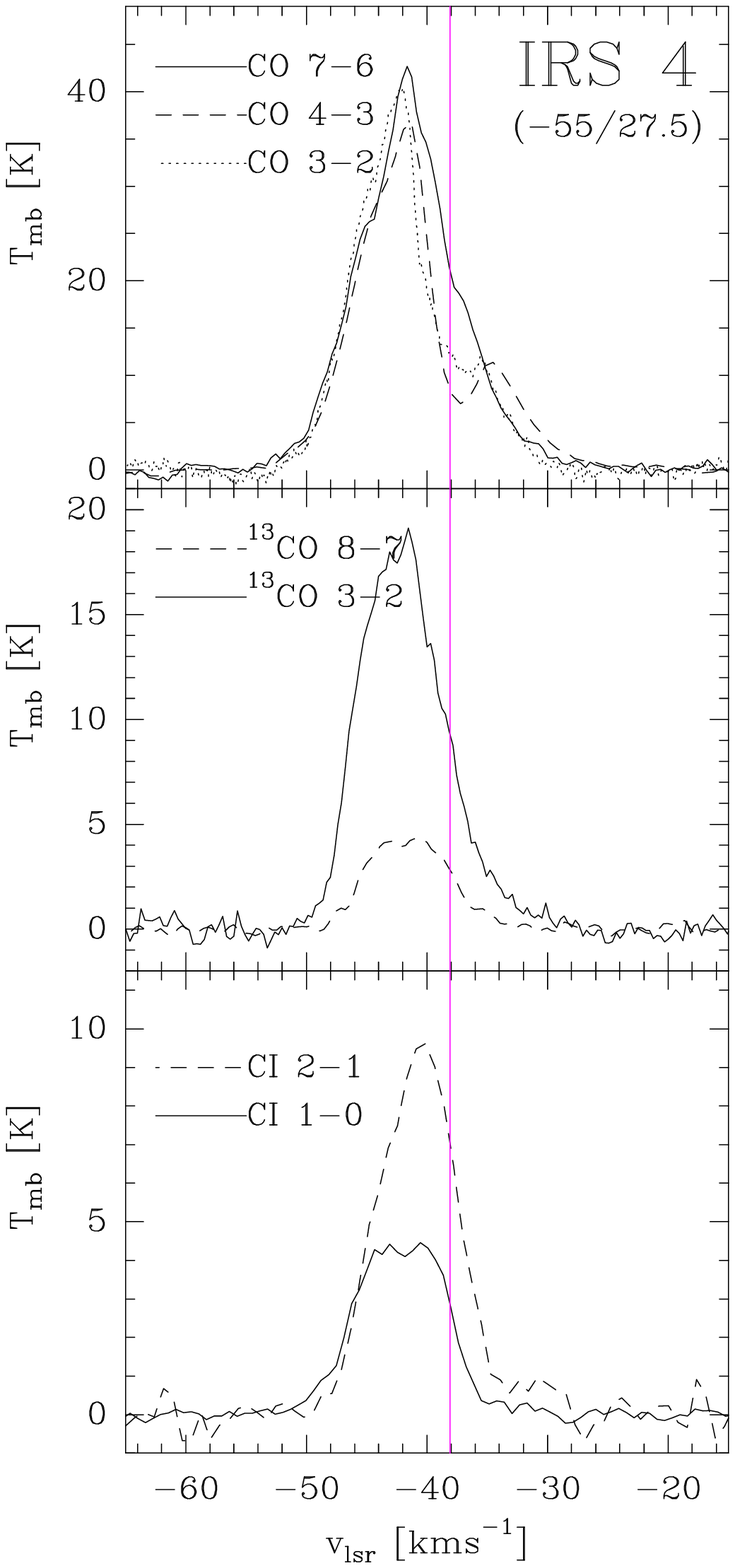}
\caption{
  Spectra at four representative positions, including IRS\,5 and
  IRS\,4.  All spectra are on a common spatial resolution of $75''$
  (0.84\,pc) and are plotted on the $T_{\rm mb}$ scale at
  0.1\,kms$^{-1}$ channel spacing.  $^{13}$CO and \CI\ lines peak at
  the velocity of a dip in the $^{12}$CO spectra, indicating that the
  dip is due to a colder foreground component at about $-38$\,$\kms$
  (denoted by a vertical line).  To estimate the line integrated
  $^{12}$CO intensity of the background source, we fitted a Gaussian
  to the line wings of the $^{12}$CO spectra.  
    Figure\,\ref{fig-irs5-gauss} gives an example.  }
\label{fig-irs4and5}
\end{figure*}

In Figure\,\ref{fig-irs4and5}, we present spectra at four positions,
IRS\,4, IRS\,5, and two positions 0.92\,pc ($85''$) to the north and
to the south of IRS\,5 where the line integrated intensities drop to
about a third of the peak intensities. In general, $^{12}$CO line
shapes show a complicated structure, espcially at IRS\,5 and 4, while
$^{13}$CO and \CI\ line shapes show little deviations from Gaussian
profiles.

At the position of IRS\,5, the $^{12}$CO 3--2 and 4--3 spectra clearly
show an absorption dip, while the $^{12}$CO 7--6 spectrum is
flat-topped. The $^{13}$CO and \CI\ spectra all peak at about the
velocity of the dip, near $-38\,\kms$. The foreground absorption in
the center region of W3\,Main is already well known from the CO
observations of \citet{hasegawa1994}. It is also seen in the water
line observed with the Odin satellite \citep{wilson2003}.  In order to
correct for self-absorption in the $^{12}$CO line profiles, we have
fitted Gaussian profiles to the line wings (Fig.\,\ref{fig-irs4and5})
to derive reconstructed integrated intensities and their ratios
(Table\,\ref{tab-wco-positions}). 
 In the following, we will only use the corrected line integrated
  $^{12}$CO intensities.

The $^{12}$CO FWHM line widths at IRS\,5 are $\sim9\,\kms$, larger
than those of the $^{13}$CO lines which are in turn larger than those
of the \CI\ lines which exhibit $\sim6\,\kms$.



\begin{figure}[th]
\centering
\includegraphics[angle=-90,width=8cm,totalheight=6cm]{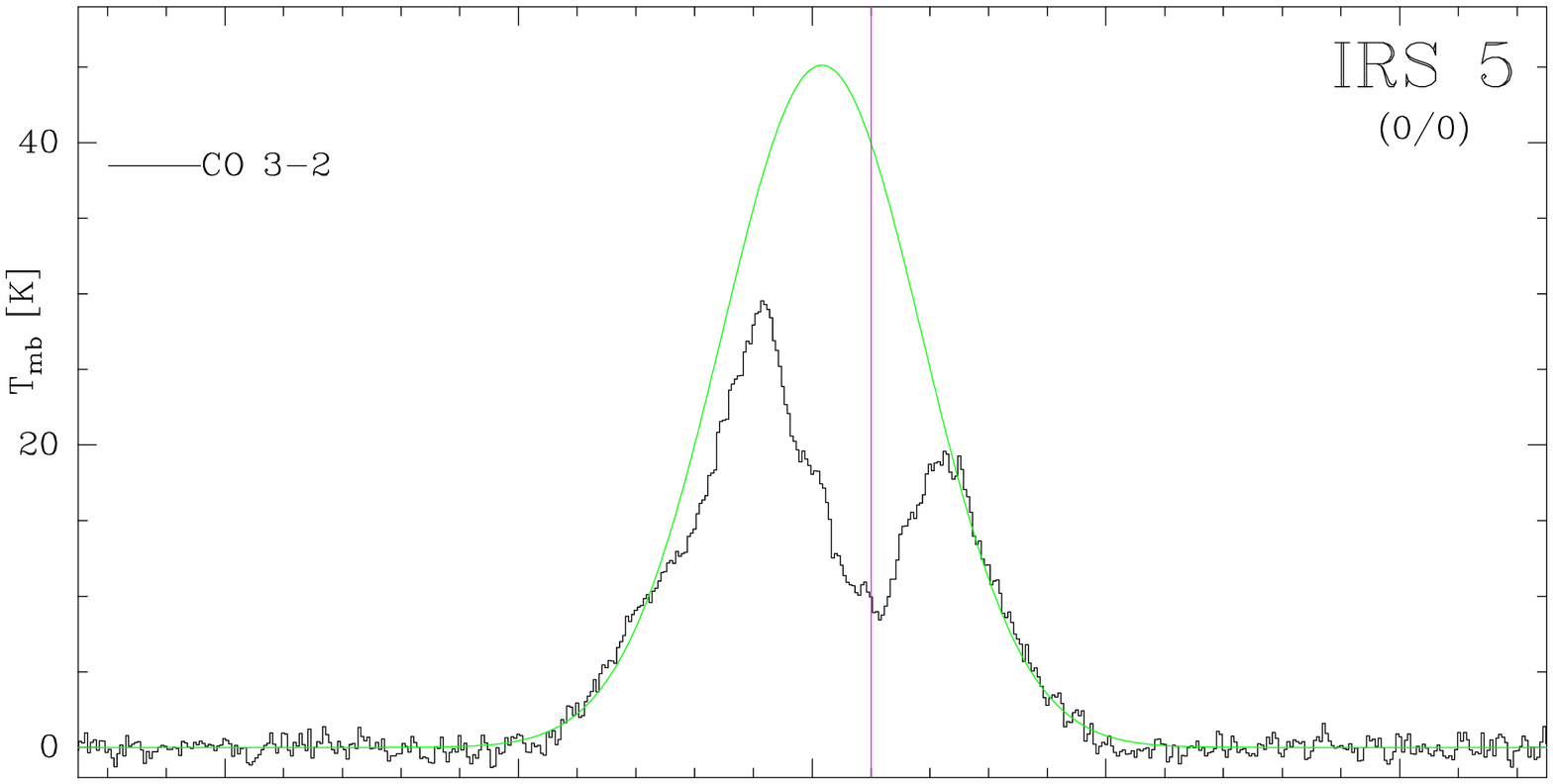}\\ 
\includegraphics[angle=-90,width=8.02cm]{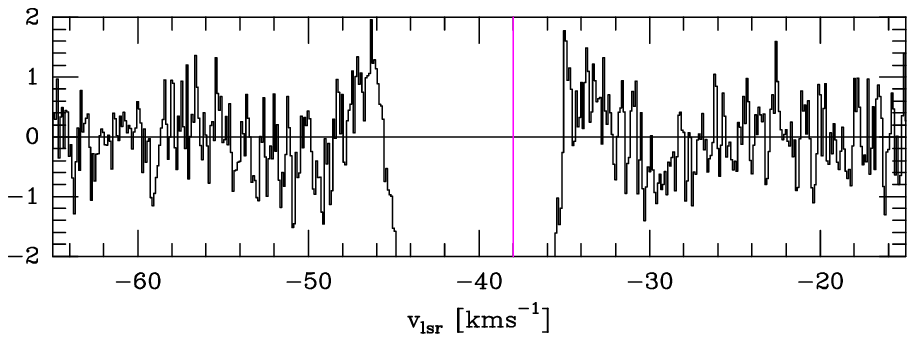}
\caption{
  The $^{12}$CO 3--2 spectrum at IRS\,5 is shown together with the
  Gaussian line profile fitted to the line wings. The ratio of fitted
  to observed line integrated intensity is 1.5. The lower box shows the
  residual of the profile fit. }
\label{fig-irs5-gauss}
\end{figure}

\subsection{Temperatures and carbon budget at four representative positions}

Observed line integrated intensities at the same angular resolution of
$75''$ and some derived line ratios are compiled in
Tables\,\ref{tab-wco-positions} and \ref{tab-wco-ratios}.  Physical
parameters at the four selected positions are listed in
Table\,\ref{tab-carbon-content}.

\subsubsection{CO line ratios, excitation temperatures, and 
beamfilling factors}
\label{sec-co-line-ratios}

 The $^{12}$CO 7--6/$^{12}$CO 4--3 line ratio of integrated
  intensities (Table\,\ref{tab-wco-ratios}), corrected for
  self-absorbed line profiles, varies between 1.0 and 1.3. This is
  consistent with optically thick emission. Assuming equal excitation
  temperatures (LTE), a ratio of $1\pm$20\% indicates temperatures of
  more than 40\,K. This is consistent within the errors with the
  excitation temperatures derived from $^{12}$CO 3--2 peak line
  temperatures in the center region and to the North, again assuming
  optically thick emission (Table\,\ref{tab-wco-ratios}).
  Excitation temperatures from CO 7--6 are $\sim10\,$K higher due to
  the Rayleigh-Jeans correction at equal peak line temperatures. 
  
  Only the ISO/LWS observations of high-$J$ CO transitions at IRS\,5
  together with an escape probability analysis
  (Chapter\,\ref{sec-esc-prob}) show that an additional gas component
  exists exhibiting temperatures of more than 140\,K.

These arguments, assume so far that radiation is emitted from a
homogeneous region. However, temperature and density gradients are
known to exist in UV illuminated clouds. We will explore their
significance when comparing the observations with PDR models in
Sect.\,\ref{chap-pdr-model}.


\begin{table*}[ht]
\caption[]{\label{tab-wco-positions}
Observed integrated intensities in $\kkms$ at the four positions in W3\,Main shown in Figure\,\ref{fig-irs4and5}.
The $^{12}$CO line intensities were corrected for self-absorption. For this reason, 
some of the integrated intensities are identical.
 The error on the integrated intensities observed with KOSMA is estimated to be $\sim15\%$.
\CII\ intensities are from \citet{howe1991}.
}
\begin{tabular}{ccccccccccccc} 
\noalign{\smallskip} \hline \hline \noalign{\smallskip}
 {$\Delta\alpha/\Delta\delta$} 
  & \multicolumn{3}{c}{$^{12}$CO} & & \multicolumn{2}{c}{$^{13}$CO} 
  & & \multicolumn{2}{c}{\CI} & & \CII \\
 \cline{2-4} \cline{6-7} \cline{9-10} \cline{12-13} 
 {$['','']$} &  
  3--2 & 4--3 & 7--6 & & 3--2 & 8--7 & & $^3$P$_1\TO^3$P$_0$ & 
  $^3$P$_2\TO^3$P$_1$ & & $^2$P$_{3/2}\TO^2$P$_{1/2}$ \\ 
\noalign{\smallskip} \hline \noalign{\smallskip}
   0,0 (IRS\,5)        & 408.4 & 408.4 & 432.4 & & 148.7 &  52.9 & &  38.3 &  73.5 & & 311.3 &  \\
 $-55.0/27.5$ (IRS\,4) & 408.4 & 408.4 & 432.4 & & 158.6 &  37.4 & &  42.5 &  78.2 & & 267.2 &  \\
   0,82.5 (North)      & 231.2 & 231.2 & 231.2 & &  63.7 &  --   & &  15.5 &  44.3 & & 148.8 &  \\
   $0,-82.5$ (South)   & 129.9 &  99.4 & 132.7 & &  60.2 &  --   & &  21.3 &  27.4 & & 311.3 &  \\
\noalign{\smallskip} \hline \noalign{\smallskip}
\end{tabular}
\end{table*}

\begin{table*}[th]
\caption[]{\label{tab-wco-ratios}
Selected ratios of observed integrated intensities in $\kkms$ at the 
four positions in W3\,Main which are shown in Figure\,\ref{fig-irs4and5}. 
 The error on the integrated intensity ratios observed with KOSMA is estimated to be $\sim20\%$.
 For comparision, we also list ratios observed in other sources by
 $^a$ \citet[][KOSMA]{schneider2003}, 
 $^b$ \citet[][HHT]{wilson2001}, 
 $^c$ \citet[][AST/RO]{zhang2001},
 $^d$ \citet[][JCMT]{petitpas1998},
 $^e$ \citet[][COBE]{fixsen1999},
 $^f$ \citet[][HHT]{stutzki1997},
 $^g$ \citet[][JCMT, Fig.2]{israel2002}, 
 $^h$ \citet[][Mt.Fuji]{yamamoto2001}, 
 $^i$ \citet[][HHT/IRAM 30m]{mao2000},
 $^k$ \citet[][IRAM PdBI]{weiss2003},
 $^m$ \citet[][AST/RO]{tieftrunk2001}.
}
\begin{tabular}{ccccccccccccccc} 
\noalign{\smallskip} \hline \hline \noalign{\smallskip}
 {$\Delta\alpha/\Delta\delta$} 
  & $^{12}$CO 3-2 & & CO 7-6 & & $^{13}$CO 3-2 & & \CI\ 2-1 & & \CI\ 1-0 & & \CII\    & & \CII\ \\
  \cline{2-2} \cline{4-4} \cline{6-6} \cline{8-8} \cline{10-10} \cline{12-12} \cline{14-14}
 {$['','']$} 
  & $^{13}$CO 3-2 & & CO 4-3 & & $^{13}$CO 8-7 & & \CI\ 1-0 & & CO 4-3   & & \CI\ 1-0 & & CO 4-3 \\ 
\noalign{\smallskip} \hline \noalign{\smallskip}
{W3\,Main:} \\
   0,0       &     2.75 & &     1.06 & &     2.81 & &     1.92 & &     0.09 & &     8.14 & &     1.12   \\
 $-55,28$    &     2.57 & &     1.06 & &     4.24 & &     1.84 & &     0.10 & &     6.29 & &     0.82   \\
   0,82.5    &     3.63 & &     1.00 & &     --   & &   (2.86) & &     0.07 & &     9.60 & &     1.11   \\
   $0,-82.5$ &     2.16 & &     1.33 & &     --   & &     1.29 & &     0.21 & &    14.62 & &     3.27   \\
\noalign{\smallskip} \hline \noalign{\smallskip}
{S106$^a$}
             & 3.0      & & 1.03--1.57   & &   & &  1.0--1.8 & & 0.12--0.38 & & 6.4 & & 0.8 \\
{Orion A}
             &          & & 1.2--3.5$^b$ & &   & &  0.83$^h$  & & \\
{Carina MC$^c$}
             &          & &              & &   & &           & & 0.14--0.45 \\
\multicolumn{2}{l}{G333.0-0.4, NGC6334A, G351.6-1.3} 
             &          & &              & &   & 0.48-0.75$^m$  \\             
{Inner Galaxy$^e$}
             &          & &              & &   & &           & & 0.31       & & 3.54 & & 4.29 \\
{Galactic Center$^e$}
             &          & &     0.11     & &   & &           & & 0.22       & & 1.38 & & 0.71 \\
{M83 Nucleus$^d$}
             &          & &              & &   & &           & & 0.14--0.47 & &  \\
{M82 Center}
             &          & & 0.28-0.36$^i$& &   & & 1.0$^f$   & & \\
{Sample of 15 galaxies$^g$}
             &          & &              & &   & &           & & 0.1--1.2 \\
{Cloverleaf}
             &          & &              & &   & & 0.54$^k$  & & \\
\noalign{\smallskip} \hline \noalign{\smallskip}
\end{tabular}
\end{table*}

\begin{table*}[th]
\caption[]{\label{tab-carbon-content}
Physical parameters at four positions in W3\,Main.
The excitation temperature $T_{\rm ex}{\rm (CO)}$ (col. (1)) is derived from $^{12}$CO 
3--2 peak temperature. 
Lower limits of the \CIw\ excitation temperature (col. (8)) are derived from the 
\CI\ 2--1/1--0 line ratio assuming a calibration error of 20\%.
For the calculation of CI column densities, it was however assumed that 
$T_{\rm ex}$(\CI) is 100\,K constant. Column densities per $75''$ beam (0.84\,pc)
and the relative abundances of the three major gas phase 
species CO, \CIw, and \CIIw, are derived assuming LTE (cf. chapter \ref{chap-rel-abundances}).
%
%
The FUV field 
$G_0$
smoothed to $75''$ is estimated from the luminosity of 
the embedded OB stars (cf. chapter \ref{chap-fuv-field}). 
}
\begin{tabular}{cccccccccr}
\noalign{\smallskip} \hline \hline \noalign{\smallskip}
  $\Delta\alpha/\Delta\delta$ & $T_{\rm ex}$(CO) & 
 $N$(CO) & $N$(\CIw) & $N$(\CIIw) & 
 \CIw:CO & \CIIw:\CIw:CO & $T_{\rm ex}$(\CIw-Ratio) &  $N$(H$_2$) & $G_0$ \\
\cline{3-5}
 $['','']$ & [K] &
 \multicolumn{3}{c}{$[10^{17}$\,cm$^{-2}]$} & & & [K] & $[10^{21}$\,cm$^{-2}]$ & 
$[10^4]$ \\
\noalign{\smallskip} \hline \noalign{\smallskip}
   $0,0$$$  &    53.&    55.2 &    5.8 &   13.8 & 10:90 & 18: 8:74 & $>122$ &  39.9 & 40.4 \\
 $-55.0,27.5$$$  & 53. &    58.7 &    6.4 &   11.8 & 10:90 & 15: 8:76 & $>108$ & 42.5 & 19.9 \\
   $0,82.5$  &    39.&     20.9 &    2.4 &    6.6 & 10:90 & 22: 8:70 & --     & 15.1 & 10.4 \\
   $0,-82.5$  & 21. & 26.2 & 3.2 & 13.8 & 11:89 & 32: 7:61 & $>54$  &  18.9 &  6.7 \\
%
%
\noalign{\smallskip} \hline \noalign{\smallskip}
\end{tabular}
\end{table*}


\subsubsection{\CI\ line ratios}
\label{chapter-ci-line-ratios}

Ratios of the two \CI\ lines vary systematically between 1.3 in the
south and 
 2.9
in the north (Table\,\ref{tab-wco-ratios}).  These
ratios indicate optically thin emission since for optically thick \CI\ 
lines we would expect line ratios of about 1 or less. The low peak
line temperatures of less than 11\,K do not reflect the kinetic
temperatures estimated above, instead they also indicate low optical
depths ($\tau<0.15$ for $T_{\rm ex}=100\,$K).

In the optically thin limit, the excitation temperature is a sensitive
function of the line ratio $R$ in $\kkms$:
$T_{\RM{ex}}=38.8\,{\RM{K}}/\ln[2.11/R]$. In Column (8) of
Table\,\ref{tab-carbon-content} we give the lower limit on $T_{\rm
  ex}$ using this formula given the observational error on the line
ratio. The error is such that only lower limits on excitation
temperatures can be derived.  Ratios above 2.11 as found at one
position north of IRS\,5 lead to unphysical negative excitation
temperatures. 
 Detailed PDR modelling by \citet[][Fig.8]{kaufman1999} confirms
  that for densities of upto $n=10^7$\,cm$^{-3}$ and FUV fields of
  upto $G_0=10^{6.5}$ the CI line ratio stays below 2.1.
  
  To exclude systematic observational errors, e.g. due to the KOSMA
  error beam at short wavelengths or due to the atmospheric
  calibration, we studied the distribution of all observed CI line
  ratios. In total 167 positions were observed, out of which 14
  positions (8.4\%) show a CI line ratio greater than 2.
  This is consistent with a true ratio of 1.5, a 20\% $1\sigma$ error,
  and a Gaussian error distribution \citep[e.g.][]{taylor1982}. Using
  the above simplifying formula for optically thin emission and equal
  $T_{\rm ex}$, a ratio of 1.5 results in a temperature of 114\,K
  which is characteristic for the average temperature of the CI
  emitting gas.  

%
%
In conclusion, carbon excitation temperatures derived from the \CI\ 
line ratios are typically 100\,K, and are higher than those derived
from $^{12}$CO. However, the sensitivity of the above ansatz on the
observational errors for high observed ratios does not allow more
quantitative conclusions.



Only few observations of the upper \CI\ line are reported so far
(Table\,\ref{tab-wco-ratios}). The 809\,GHz line was first detected by
\citet{jaffe1985} and later observed by \citet{genzel1988} in M\,17
and W\,51. The COBE satellite observed the Milky Way at $8^\circ$
resolution and found low line ratios \citep{fixsen1999} of 0.31 in the
inner Galactic disk, corresponding to a kinetic temperatures of
$>20$\,K. Line ratios observed in Orion\,A (OMC1)
\citet{yamamoto2001} using the Mt.Fuji telescope indicate excitation
temperatures covering the range of 52 to 62\,K and more.
\citet{zmuidzinas1988} detected the upper \CI\ line in several
Galactic star forming regions with the KAO, and derived excitation
temperatures between 25\,K and 77\,K from the \CI\ 2--1/1--0 line
ratios. They derived a carbon excitation temperature of 25\,K from
observations of W3. This is in contradition to our result which is
however based on simultaneous observations with the same instrument.
Moreove data were smoothed to the same spatial resolution thus
reducing possible calibration errors.

\subsection{The case for hot gas at IRS\,5}

\citet{helmich1994} found rotational temperatures of SO$_2$, CH$_3$OH,
and H$_2$CO ranging between 64 and 187\,K when using the JCMT with
high spatial resolutions.  \citet{vandertak2000} find rotational
temperatures of CH$_3$OH and H$_2$CO of 73 and 78\,K.

High-$J$ lines of CO and $^{13}$CO $v=0-1$ (4.7\,$\mu$m) absorption
measurements towards IRS\,5 with a subarcsecond sized beam given by
the size of the IR continuum background source \citep[][ at the
CFHT]{mitchell1990} reveal a hot gas component at 577\,K comprising
$\sim50\%$ of the total gas column density per beam
 at densities exceeding $10^7$\,cm$^{-3}$.  
A colder component at $\sim66$\,K comprises the other half.

Solid CO is also detected towards IRS\,5
\citep{sandford1988,lacy1984,geballe1986}, but contributes only
0.2\,\% to the total CO column density \citep{mitchell1990}. ISO-SWS
observations of the $^{13}$CO$_2$ ice band profile \citep{boogert2000}
towards IRS\,5 show indications of heated ice, consistent with the
above results.

Ammonia observations at resolutions of $40''$ were used by
\citet{tieftrunk1998a} to derive kinetic temperatures at the positions
of IRS\,5 and 4. While the kinetic temperature at IRS\,4 is estimated
to be 48\,K in excellent agreement with the excitation temperature we
have derived from $^{12}$CO 3--2, the kinetic temperature at IRS\,5
was estimated to be 150\,K.

In conclusion, observations indicate a strong temperature gradient
peaking at IRS\,5 at upto 600\,K.  The hot gas component has a low
beam filling factor while the colder gas component is more widespread.
Variations of optical depths and chemistry are also at play.

\subsubsection{Relative carbon abundances}
\label{chap-rel-abundances}

Table\,\ref{tab-carbon-content} lists the relative abundances of the
major carbon bearing species in the gas phase: \CIIw, \CIw, and CO.
These are derived from column densities assuming optical thin
$^{13}$CO 3--2, \CI, and \CII\ emission and LTE. For the \CIw\ column
density we assumed an excitation temperature of 100\,K at all
positions. Note that the column density is not temperature sensitive:
an increase of $T_{\rm ex}$ by a factor of 2 only results in a 10\%
increase of the column density. \CIIw\ column densities are independent
of the excitation temperature in the thermalized limit and for
temperatures much higher than 91\,K \citep[see e.g.][]{schneider2003}.

The total H$_2$ column density is derived from the $^{13}$CO 3-2
integrated intensity via the CO column density assuming a
$^{12}$CO/$^{13}$CO abundance of 65 \citep{langer_penzias1990} and a
H$_2$/CO abundance of $8\,10^{-5}$.

In W3\,Main, the relative abundances of \CIw/CO are rather constant:
the ratio is $\sim0.11$ at all four positions.  Similar ratios of
$\sim0.1$ are found by \citet{zmuidzinas1988} in a study of seven
Galactic star forming regions. \citet{plume2000} derive similar CI/CO
ratios in the Orion A cloud, but observed an increase towards the
cloud edges to values of $\sim0.5$.  \citet{schneider2003} find low
ratios in S\,106 varying between 0.09 and down to 0.03.
\citet{howe2000} find ratios of about 0.4 in M17SW.  \citet{gerin1998}
find ratios of between 0.1 and 0.7 in the photodissociation region
associated with the reflection nebula NGC\,7023.  The quiescent
translucent and thus less well shielded cloud MCLD\,123.5$+$24.9
exhibits high \CIw/CO ratios of between 0.2 and 1.1 \citep{bensch2003}.

Back to W3\,Main: The \CIw\ abundance relative to both other two major
gas-phase carbon species is constant at $\sim8\%$. The \CIIw\ 
abundances is high in the south (32\%) and lower at the other
positions.  In other words, the relative CO abundance is greater than
60\% throughout.  Similar fractional abundances of \CIIw:\CIw:CO as at
W3\,Main-South are seen in IC\,63 \citep{jansen1996}: 37:7:56, still
lower abundances of CO are seen in NGC\,2024 (Orion\,B)
\citep{jaffe1995}: 40:10:50. \citet[][]{brooks2003} found fractional
abundances for \CIIw:\CIw:CO of 68:15:16 in $\eta$Carina.  On the
other hand, CO abundances are much higher in S106
\citep{schneider2003} at all positions studied ($>86\%$), dwarfing the
contributions from \CIIw\ and \CIw. In S106, \CIIw\ contributes 8\% to
the carbon content at the position of the exciting star S106\,IR, even
less at all other positions. In contrast, the \CIIw\ fractional
abundance is higher all over W3\,Main, probably due to the much larger
number of embedded OB stars.

\subsection{Total H$_2$ column densities and masses}
\label{sec-totaln}

Total H$_2$ column densities per $75''$ beam derived from $^{13}$CO
vary between 1.5 and $4.3\,10^{22}$\,cm$^{-2}$ at the four positions
(Table\,\ref{tab-carbon-content}). This agrees within a factor of 2
with total column densities derived from C$^{18}$O at $40''$
resolution by \citet{tieftrunk1998a}. 

The total mass per $75''$ beam from $^{13}$CO at IRS\,5 is
481\,M$_\odot$.  Here, we assume a molecular mass per hydrogen atom of
1.36.

\section{The physical structure of W3\,Main}
\subsection{Modelling the spatial structure of the FUV field}
\label{chap-fuv-field}

\begin{figure}[htb]
\centering
\includegraphics[angle=-90,totalheight=6.5cm]{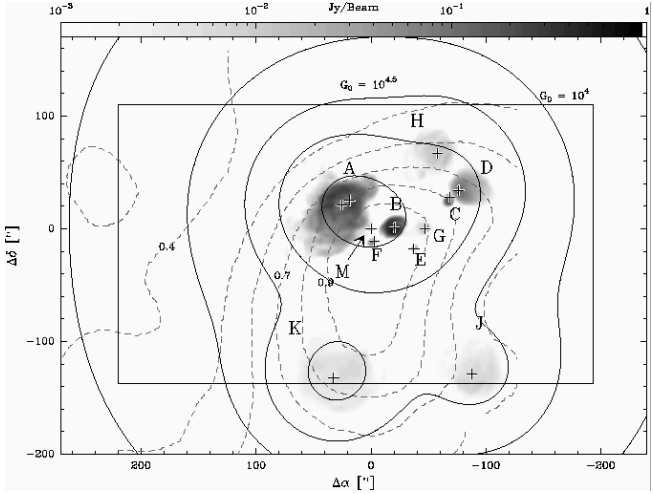}
\caption{Overlay of UV field, \CII\ emission, and 6\,cm radio continuum.
  {\bf Grey-scale:} Image of the VLA 6\,cm radio continuum
  \citep{tieftrunk1997}. The box denotes the region observed with
  KOSMA.  {\bf Solid contours:} Contours of the calculated 
  spatial structure of the FUV field smoothed to $75''$
    resolution.
  The positions of the OB stars are marked with a $+$.  {\bf Dashed
    contours:} extended [\ion{C}{ii}] emission observed by
  \citet{howe1991} with the KAO at $80''$ resolution.  Contour levels
  range from 40\% to 90\% in steps of 10\% of the peak intensity of
  $314\,\kkms$.
\label{fig-cii}
}
\end{figure}

\citet{tieftrunk1997} presented a multi-frequency VLA study of the
radio contiuum sources in W3\,Main. This region exhibits the diffuse
regions W3\,H, J, and K, the compact regions W3\,A, B, and D, the
ultracompact regions W3\,C, E, F, and G, and hypercompact regions with
diameters of less than $1''$ ($\sim0.01$\,pc) W3\,Ma-g and Ca which
probably correspond to the youngest evolutionary stage of an \HII\ 
region.  Figure\,\ref{fig-cii} shows the positions of the 17 OB stars
exciting the \HII\ regions
\citep[Table\,\ref{tab-obstars}][]{tieftrunk1997}.  The most luminous
source is IRS\,2, identified as an O5 star. The luminosity of IRS\,5
corresponds to a cluster of $\sim7$ B0-type stars
\citep{claussen1994}. We use the spectral types of the exciting OB
stars derived by \citet{tieftrunk1997} to calculate the spatial
large-scale variation of the FUV field heating the molecular cloud.

The FUV field of W3\,Main was modelled in two dimensions by summing
the contributions from all sources at each pixel of a regular grid.
For all sources, we assume that the positions inside the cloud are
given by their projected distances. The FUV field was then derived
assuming that the stars are ZAMS stars at the effective temperature
corresponding to their spectral type \citep{panagia1973}. We furtheron
assume only geometrical dilution of the flux, i.e. we neglect
scattering and absorption of UV by dust and clumps, thus deriving
upper limits only. 
%

Due to the many bright embedded sources, the resulting FUV field,
smoothed to $75''$ resolution, varies by a factor of 100, dropping
from 
 $G_0=10^{5.5}$ in the center to $10^{3.5}$ at the edges of the
  mapped region. The Habing-field $G_0$ used throughout this paper is
  in units of $1.6\,10^{-3}$\,erg s$^{-1}$cm$^{-2}$ \citep{habing1968}.
  See \citet{draine1996} for a discussion of other measures of the
  average interstellar radiation field.
%

 The total FIR luminosity of W3 given in the literature e.g.  by
  \citet{thronson1980} is $1\,10^6$\,L$_\odot$. Solely the luminosity
  of the O5 star powering IRS\,2 accounts already for 68\% of this
  (cf. Table\,\ref{tab-obstars}).  The total luminosity of all OB
  stars is $1.5\,10^6$ L$_\odot$.  This indicates that the primary
  source of the FIR continuum is the reemitted stellar radiation field
  of the OB stars and that some of the stellar photons escape without
  impinging on the clouds.  Additional heating sources like e.g.
  outflow shocks, stellar winds, or the embedded low mass objects are
  not needed to explain the observed total FIR luminosity.  

\begin{table}
      \caption[]{\HII\ regions and exciting OB stars which were used
to construct the two-dimensional FUV-field shown as drawn contours
in Figure\,\ref{fig-cii}. The spectral type (col. (2)) 
is an average from the compilation of
\citet{tieftrunk1997}. Total luminosities (col. (4)) assuming ZAMS stars are from 
\citet{panagia1973}. Column\,(5) gives the corresponding names of
infrared sources \citep{wynn1972,richardson1989}.
      }\label{tab-obstars}
\begin{tabular}{llcrl}
\noalign{\smallskip} \hline \hline \noalign{\smallskip}
\HII\  & Spectral & Position & $L$ & Remark\\
Region & Type     & ($'',''$) & $10^4$L$_\odot$ \\
\hline
A   & O5   & $(25,21)   $ & $68.0$   & IRS\,2\\
B   & O6.5 & $(-20,1)   $ & $15.0$  & IRS\,3a\\
K & O6.5  & $(32,-132)$    & 15.0 \\ 
J & O8    & $(-87, -129)$  & 6.5 \\ 
D   & O8.5 & $(-76,34)  $ & $5.4$  & IRS\,10\\
Aa  & O8.5 & $(18,25)   $ & $5.4$  & IRS\,2a\\
H   & O8.5 & $(-58,67)  $ & $5.4$  & \\
F   & O9.5 & $(-3,-11)  $ & $3.8$  & IRS\,7\\
C   & O9.5 & $(-72,27)  $ & $3.8$  & IRS\,4\\
G   & B0   & $(-47,0)   $ & $2.5$ & \\
E   & B0   & $(-37,-18) $ & $2.5$  & \\
Ma-g&B0    & $(0,0)     $ & $7\times2.5$  & IRS\,5\\
Ca  & B1   & $(-69,28)  $ & $0.5$ & IRS\,4\\
\hline
Sum &      &              & 151.0 & \\
\noalign{\smallskip} \hline \noalign{\smallskip} 
\end{tabular}
\end{table}

\subsection{A map of \CII\ emission}\label{ciidata}

Figure\,\ref{fig-cii} shows the extended [\ion{C}{ii}]
$^2$P$_{3/2}\TO^2$P$_{1/2}$ 158~$\mu\RM{m}$ emission
\citep[][]{howe1991} in contours at a resolution of $80''$.
%
%
The \CII\ emitting region shows a north-south extension and extends
over all the \HII\ regions
which have been identified in radio continuum emission studies
\citep{tieftrunk1997}. Interestingly, the north-south orientation is
not seen in the atomic carbon maps nor in the CO maps
(Figures\,\ref{midj}). Neither is it seen in the FIR continuum at
158\,$\mu$m observed by \citet{howe1991} which shows a circular
symmetry peaking at IRS\,5. This could mean that \CII\ does
not solely trace PDRs, i.e. the molecular cloud surfaces.  A fraction
of \CII\ emission may stem from the \HII\ regions and the diffuse
outer halo. In the following paragraph, we estimate the contribution
by the \HII\ regions in W3\,Main.

In addition, the modelled FUV field does not trace well the observed
distribution of \CII. The former shows a peak at IRS\,5 while the
latter shows a north-south orientated ridge of constant \CII\ 
emission.  This mismatch is surprising since \CII\ is thought to be an
accurate tracer of the FUV field.
%
\citet{howe1991} showed that a homogeneous cloud model cannot
reconcile this particular discrepancy. Rather a clumpy cloud model has
to be invoked with dense clumps immersed in a dilute interclump
medium, allowing the FUV radiation to deeply penetrate into the cloud.
Here, we try to improve on this analysis by using the new observations
of \CI\ and CO.
%
%
 
\subsection{\CII\ emission from the ionized medium}
\label{sec-cii-hii}

\CII\ emission does partly stem from the \HII\ regions. However, the
contribution of \CII\ emission from \HII\ regions to the observed
\CII\ emission is in most cases found to be weak. 
We have calculated the fraction of \CII\ emission from the \HII\ 
regions in W3\,Main using the emission measures and electron densities
compiled by \citet{tieftrunk1997} and the formalism of
\citet[][]{russell1981} \citep[cf.][]{crawford1985,nikola2001}.  We
found that the contribution is very low, varying between 0.1 and
0.7\%.

This result appears to be valid also on much larger scales. Using the
COBE Milky Way survey, \citet{petuchowski1993} find that \HII\ regions
contribute only $\sim1\,$\% of the total luminosity of the Galaxy in
the \CII\ line.


\begin{figure}[htb]
\centering
\includegraphics[angle=-90,width=\linewidth]{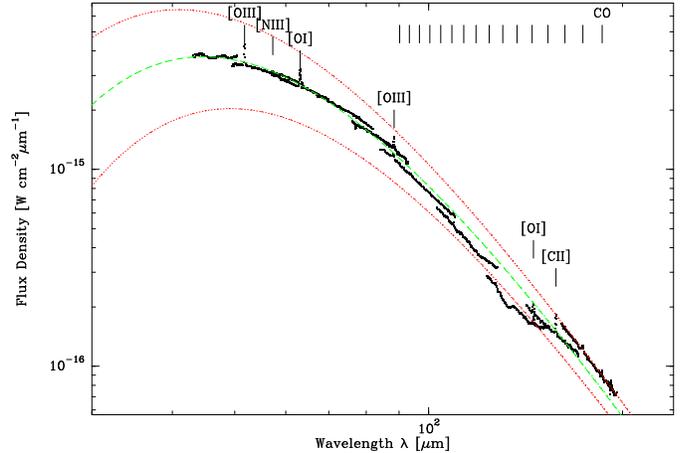}
\caption{ISO/LWS spectrum of the IRS\,5 position in W3\,Main 
  The overlapping LWS bands are clearly discernable. The dashed line 
  is the isothermal grey body fit which results in $T_{\rm
    dust}=53$\,K and $\beta=1.2$ for $A_V=40$\,mag. For comparison,
  the drawn lines show the grey body for $58\,$K and $47\,$K dust
  temperature at the same $\beta$ value. The positions of several
  atomic fine structure and high-$J$ CO lines are marked.
\label{fig-bb-irs5}
}
\end{figure}

\begin{figure}[htb]
\centering
\includegraphics[angle=-90,width=\linewidth]{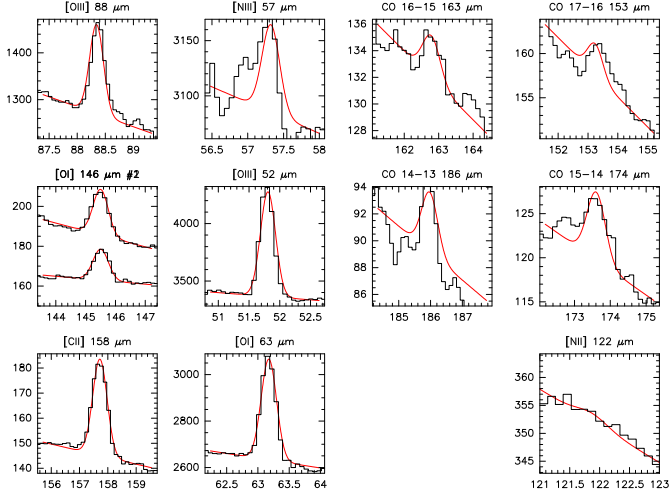}
\caption{Individual ISO/LWS spectra at IRS\,5/W3\,Main (cf. Fig.\,\ref{fig-bb-irs5}). 
  Drawn lines show the fitted instrumental line profiles. Integrated
  line intensities are listed in Table\,\ref{tab-isodata}.
\label{fig-iso-lws-lines}
}
\end{figure}
 
\begin{table}
\begin{center}
\caption{\label{tab-isodata}
  ISO/LWS integrated intensities of the FIR lines and the FIR
    continuum at IRS\,5.  Integrated intensities are listed in units
  of [$10^{-3}$erg\,s$^{-1}$\,cm$^{-2}$\,sr$^{-1}$] and were derived
  from the measured flux per beam using the beam sizes (col. (4)) and
  corrections for extended source emission given in Table 4.10 in the
  ISO-LWS Handbook v. 1.2. The listed errors only reflect the line
  fitting uncertainty.
  The intensity of the \OI($145\,\mu$m) line is the average of 
  two LWS detectors. The error is in this case the deviation from the 
  individual fluxes.
}
\begin{tabular}[h]{llrlr}
\noalign{\smallskip} \hline \hline \noalign{\smallskip}
&IRS\,5      & $\lambda$  & Integr.Intensity & HPBW \\
&            & [$\mu$m]   &            & [$''$]\\
\noalign{\smallskip} \hline \noalign{\smallskip}
& CO 14--13  & 186. &  0.21 $\pm$  0.03 & 66. \\
& CO 15--14  & 174. &  0.30 $\pm$  0.03 & 66. \\
& CO 16--15  & 163. &  0.08 $\pm$  0.03 & 66. \\
& CO 17--16  & 153. &  0.13 $\pm$  0.03 & 68. \\
%
%
\hline
& \CII       & 158. &  1.54 $\pm$  0.03 & 68. \\
& \OI        & 145. &  0.77 $\pm$  0.13 & 68. \\
& \NII & 122. &  0.06 $\pm$  0.06 & 78. \\
& \OIII      &  88. &  3.8 $\pm$  0.2 & 76. \\
& \OI        &  63. &  8.3 $\pm$  0.3 & 86. \\
& \NIII      &  57. &  1.7 $\pm$  0.3 & 84. \\
& \OIII      &  52. & 19 $\pm$  1 & 84. \\
\hline 
& FIR-Continuum & & 13500 & \\
\hline
\end{tabular}
\end{center}
\end{table}

\subsection{ISO/LWS data of IRS\,5/W3\,Main}

We retrieved the FIR ISO/LWS data at IRS\,5 from the data archive and
reanalyzed it as described in Sect.\,\ref{sec-iso-observations}.

\subsubsection{Grey body fit}
\label{sec-grey-body-fit}

The far-infrared continuum at IRS\,5 is modelled using the ISO/LWS
data.  We assumed an isothermal grey body spectrum $F_\lambda = \Omega
B_\lambda(T_{\rm dust}) (1-\exp^{-\tau_{\rm dust}})$, where the solid
angle $\Omega$ was fixed to correspond to the average LWS HPBW of
$\sim75''$, $B_\lambda(T_{\rm dust})$ is the Planck function at
temperature $T_{\rm dust}$, and the dust opacity is parametrized by
$\tau_{\rm dust}=\tau_V(\lambda/\lambda_V)^{-\beta}$.  The total H$_2$
column density derived in Sect.\ref{sec-totaln} corresponds to an
optical extinction of $A_V=40$\,mag$=1.086\,\tau_V$ \citep{bohlin1978}
and was held fixed as well. A least squares fit of the dust
temperature and the spectral index results in $T_{\rm dust}=53\pm1$\,K
and $\beta=1.21\pm0.02$ (Fig.\,\ref{fig-bb-irs5}).  The total
brightness, obtained by integrating over all wavelengths, is
 $2.1\,10^{-13}$\,Wcm$^{-2}$beam$^{-1}$ or 13.5\,erg s$^{-1}$ cm$^{-2}$
sr$^{-1}$. The total luminosity within the beam is $3.4\,10^5$\,L$_\odot$.
%
%
Comparison with Table\,\ref{tab-obstars} indicates that all of the FIR
luminosity is powered by the O5 star associated with IRS\,2
(Table\,\ref{tab-obstars}). 

 Assuming $I_{\rm FIR} = 2\times1.6\,10^{-3} (4\pi)^{-1} G_0$
  \,erg s$^{-1}$cm$^{-2}$sr$^{-1}$ \citep{kaufman1999}, i.e. assuming
  that the stellar FUV flux is solely responsible for the FIR
  continuum, we estimate the FUV field at IRS\,5 to be
  $G_0=5.3\,10^4$. This is only 13\% of the FUV field estimated solely
  from the spectral types of the OB stars: $G_0=4\,10^5$
  (Table\,\ref{tab-carbon-content}) which may again indicate that a
  part of the stellar photons escape the cloud without heating the
  clouds. This issue will be discussed again in
  Chapter\,\ref{sec-total-fluxes} when comparing with the results of
  the PDR modelling.

To improve on the analysis of the ISO/LWS continuum, a better
calibration of the LWS bands would be needed and the variation of LWS
beam sizes ($55''-90''$) would need to be taken into account.


\citet{oldham1994} derive slightly higher values of $T_{\rm
  dust}=68$\,K and $\beta=1.4$ from fitting a single-temperature grey
body spectrum to FIR data at higher angular resolutions of $<27''$,
indicating again an increase in temperature with smaller beam.  They
also fit an optical extinction of $A_V=281$\,mag for a source diameter
of $13''$. Comparing this $A_V$ with the $A_V$ we derive within a
$75''$ beam, indicates a source size of $28''$.

\subsubsection{Atomic fine structure lines}

Integrated intensities of the transitions detected with ISO/LWS at IRS\,5
are listed in Table\,\ref{tab-isodata}.

\paragraph{Tracers of PDRs.}
The lines of \CIIw, \OIw, and CO are prominent tracers of the PDRs at
the surfaces of the molecular clouds. Of all FIR lines tracing the
neutral medium, the $63\mu$m line of \OIw\ is the major coolant at IRS\,5
being a factor of 4.1 stronger than the \CII\ line. Moreover, this
\OI\ transition may be optically thick and thus its intensity strongly
reduced by foreground absorption. This is indirectly confirmed by
detailed PDR modelling described in the Sect.\,\ref{chap-pdr-model}.



The \CII\ integrated intensity detected with ISO/LWS is 70\% of the
intensity measured by \citet{howe1991} with the KAO and thus in
agreement within the expected calibration error.

%
%
%

\paragraph{Tracers of the ionized gas.}

The ions \NIIw, \OIIIw, \NIIIw, have ionization potentials larger than
13.6\,eV and their fine structure lines therefore trace the ionized
medium.  Although, the \NII\ line at $122\,\mu$m is not detected, the
lower limit of the $\CII/\NII$ line flux ratio can be estimated to at
least 27. A ratio of 9 \citep[cf.][]{nikola2001} would be expected for
\CII\ being emitted entirely from the ionized gas. This indicates that
at most 33\% of the \CII\ emission stems from the diffuse ionized
medium.  A more stringent estimate is derived using the properties of
the \HII\ regions in W3\,Main as discussed in
Sect.\,\ref{sec-cii-hii}.



\begin{figure}[htb]
\centering
\includegraphics[angle=0,width=0.8\linewidth]{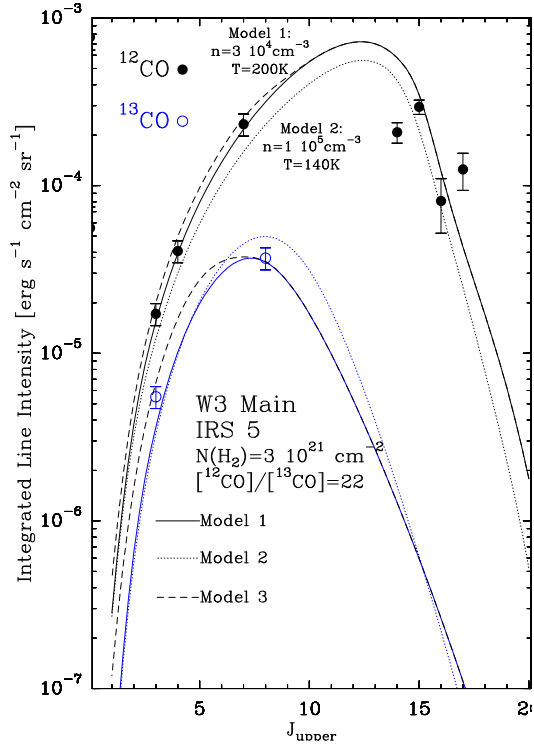}
\caption{Integrated CO intensities as a function of upper rotational quantum number 
  $J$ for IRS\,5 in W3\,Main and CO cooling curves produced by a
  non-LTE radiative transfer model. Rotational transitions upto $J=8$
  were observed with KOSMA for which we plot the 15\% uncertainty. The
  high-$J$ CO transitions upto $J=17$ were observed with ISO/LWS. We
  plot the uncertainties listed in Table\,\ref{tab-isodata}.  Model 3
  is a two-component model adding the contributions of the warm gas
  (model 1) to a cold component at $T_{\rm kin}=40\,$K,
  $n=10^3$\,\cmcub\ containing 90\% of the mass.
\label{fig-co-cooling-irs5}
}
\end{figure}

\subsubsection{Carbon monoxide emission lines}
\label{sec-esc-prob}

Four high-lying rotational lines of gas phase CO were detected with
ISO/LWS at IRS\,5. These lines trace the hot and dense gas, upper
level energies lie above 500\,K and critical densities are about
$10^7$\,\cmcub\ and higher.  To better constrain the physical
parameters, we used a non-LTE escape probability radiative transfer
model \citep{stutzki_winnewisser1985} assuming a CO abundance of
$8\,10^{-5}$ and using cross sections from \citet{schinke1985}.
Figure\,\ref{fig-co-cooling-irs5} shows the CO and $^{13}$CO line
integrated intensities observed with KOSMA and ISO/LWS.  


Although the model simplifies the source structure by neglecting
temperature and density gradients, and varying beam filling factors
are not considered, common solutions can be found to fit all 9
observed line intensities.  The minimum column density to reproduce
the observed CO fluxes is $N$(H$_2$)$=3\,10^{21}$\,\cmsq, only about
10\% of the total H$_2$ column density at IRS\,5
(Table\,\ref{tab-carbon-content}).
 (We assumed line widths of 9\,kms$^{-1}$ for the $^{12}$CO lines
  and 6\,kms$^{-1}$ for the $^{13}$CO lines.) 
The column density is largely determined by the mid-$J$ CO
transitions, rather independent of the kinetic temperatures and
densities. The kinetic temperatures need to be high to reproduce the
high-$J$ CO transitions detected with ISO/LWS while the ISO
transitions are subthermally excited. However, two solutions exist
 which bracket the parameter space of acceptable models: 
{\bf 1.} a high gas temperature of 200\,K and rather low density of
$3\,10^4$\cmcub\ and {\bf 2.} a lower gas temperature of 140\,K and a
higher gas density of $10^5$\,\cmcub. Both solutions indicate a
pressure ($n\,T$) of $(1-5)\,10^{7}$\,Kcm$^{-3}$. The CO 15--14
transition is only slightly stronger than the CO 7--6 transition, the
ratio of both intensities is 1.29. On the other hand, the intensity
ratio of CO 15--14 over CO 1--0, derived from the model results, is
about 500.  \citet{tieftrunk1998a} derived a kinetic temperature of
150\,K at IRS\,5 from ammonia observations, consistent with our
results.

However, only about 10\% of the gas is at these high temperatures and
densities. A second phase must exist containing the bulk of the gas
mass and at significantly lower temperatures. We model this phase
assuming $N$(H$_2$)$=3\,10^{22}$\,\cmsq, $T_{\rm kin}=40\,$K,
$n=10^3$\,\cmcub. The dashed curve in
Figure\,\ref{fig-co-cooling-irs5} shows the combined flux (model 3) of
the warm phase (model 1) and the cold phase. The cold phase hardly
contributes to the observed cooling fluxes of the mid-$J$ and high-$J$
CO transitions, although it contains 90\% of the gas mass.

 A [CO]/[$^{13}$CO] abundance ratio of 22 is needed to fit
  simultaneously the observed $^{13}$CO lines. The $^{13}$CO 8--7 line
  in particular is crucial in determining the column density and
  abundance ratio. The [CO]/[$^{13}$CO] abundance is significantly
  lower than the ratio of 89 found by \citet{mitchell1990} at
  sub-arcsecond resolutions, tracing the immediate hot and dense
  environment in front of IRS\,5. The latter ratio is consistent with
  the canonical interstellar abundance of 65
  \citep{langer_penzias1990}.  It is unlikely that $^{13}$CO
  fractionation plays a major role at W3/IRS\,5 since the kinetic
  temperatures are much higher than 36\,K, the energy gained in
  substituting $^{12}$C by $^{13}$C in CO.  The low [CO]/[$^{13}$CO]
  abundance ratio derived from the escape probability model may
  therefore indicate that a more detailed source model is needed,
  including density and kinetic temperature gradients along the line
  of sight. Indeed, in a more realistic model as the PDR model, the
  15--14 and 4--3 CO lines are not produced by gas at the same
  temperature and the same density.
%
%

A similar analysis as we did for IRS\,5 combining submillimeter and
far-infrared CO lines was conducted by \cite{harris1987,stutzki1988},
and \citet{stacey1993} at FIR peaks of the Galactic star forming
regions M\,17, S\,106, and
in the Orion molecular cloud. The derived minimum column densities of
the warm gas and the kinetic temperatures are very similar
($N$(CO)$=(0.5-6)\,10^{18}\,$\cmsq, $T_{\rm kin}=100-450$\,K,
$(1-4)\,10^4\,$\cmcub) to the solution we find in W3\,IRS\,5.  These
conditions appear to be typical for Galactic molecular clouds in the
vicinity of massive star formation.  Similar conditions of the warm
molecular gas, but on a much larger scale of 180\,pc are found in the
starburst nucleus of NGC\,253 by analyzing CO cooling curves
\citep{bradford2004zermatt}.

\subsubsection{Line cooling from \CII, \CI, CO, and \OI}
\label{sec-line-cooling}

 The ratio of \CII\ intensity over the FIR continuum within the
  ISO/LWS beam of $\sim75''$ at IRS\,5 is $1.1\,10^{-4}$ (cf.
  Table\,\ref{tab-isodata}). The same ratio averaged over the entire
  mapped region is a factor 14 higher, i.e.  $1.5\,10^{-3}$
  \citep{howe1991}.  Similar values of about $10^{-4}$ as found
  locally at W3\,IRS\,5 have also been found in other Galactic star
  forming regions like Orion\,A, M\,17, W\,51, or DR\,21 \citep[see
  Table\,5 in][]{stacey1991}. On the other hand, the COBE large scale
  low resolution map of the Milky Way shows a higher and rather
  constant ratio with little scatter between $50\,10^{-4}$ and
  $170\,10^{-4}$ \citep{fixsen1999}.  \CII/FIR ratio ratios of upto
  0.01 have been found in other Galactic clouds \citep{stacey1991}
  using KAO and in external galaxies \citep{stacey1991,malhotra2001}.
  Based on ISO observations, \citet{malhotra2001} found typical
  \CII/FIR ratios for normal galaxies (excluding AGNs) in the range
  0.01 to 0.001.  However, there are again significant exceptions: two
  warm and active galaxies show ratios of less than $2\,10^{-4}$,
  reminicent of the ratio we find at IRS\,5. A significant \CII\ 
  deficiency has also been found in ULIRGs \citep{luhman1998}. A
  number of explanations are discussed in the literature. One
  possibility is that increased $G_0/n$ ratios lead to increased
  positive grain charge which in turn leads to less efficient heating
  of the gas by photoelectrons from dust grains. See e.g.
  \citet{kaufman1999} for a discussion.
%
%


The sum of all observed CO intensities is $1.0\,10^{-3}$erg
s$^{-1}$cm$^{-2}$ sr$^{-1}$, only a fifth of the total modelled CO
line cooling for $J$ upto 20 (Model\,1). 
 The total CO flux is only weakly influenced by self-absorption,
  since we have corrected for this effect to first order, and
  secondly, the CO cooling is mainly produced by the highly-excited
  lines which are less affected by self absorption.
%
%
%
  The importance of the unobserved transitions between $J=8$ and
  $J=13$ to the total cooling is evident from
  Figure\,\ref{fig-co-cooling-irs5} and from the escape probability
  analysis which shows that the peaks of the best fitting modelled
  cooling curves both lie near $J=12$.
  
  The total CO cooling by far exceeds the gas cooling by the two
  carbon lines: 
$\Lambda_{\rm CO}/\Lambda_{\rm C}=123$. The two \OI\ lines are more
important: $\Lambda_{\rm O}/\Lambda_{\rm CO}=1.9$, $\Lambda_{\rm
  O}/\Lambda_{\rm CII}=6$. The relative contributions to the total gas
cooling are $\Lambda_{\rm CO}=31.6\%$, $\Lambda_{\rm O}=58.7\%$,
$\Lambda_{\rm CII}=9.7\%$, $\Lambda_{\rm C}=0.3\%$. The relative
contribution of \OI\ may be even stronger due to high optical depth
and foreground absorption of the \OI($63\,\mu$m) line. The total gas
cooling flux is $15.5\,10^{-3}$erg s$^{-1}$cm$^{-2}$ sr$^{-1}$ and
thus 
 0.1\%
of the dust cooling FIR flux.




\begin{figure*}[htb]
\centering
\includegraphics[angle=0,totalheight=6cm]{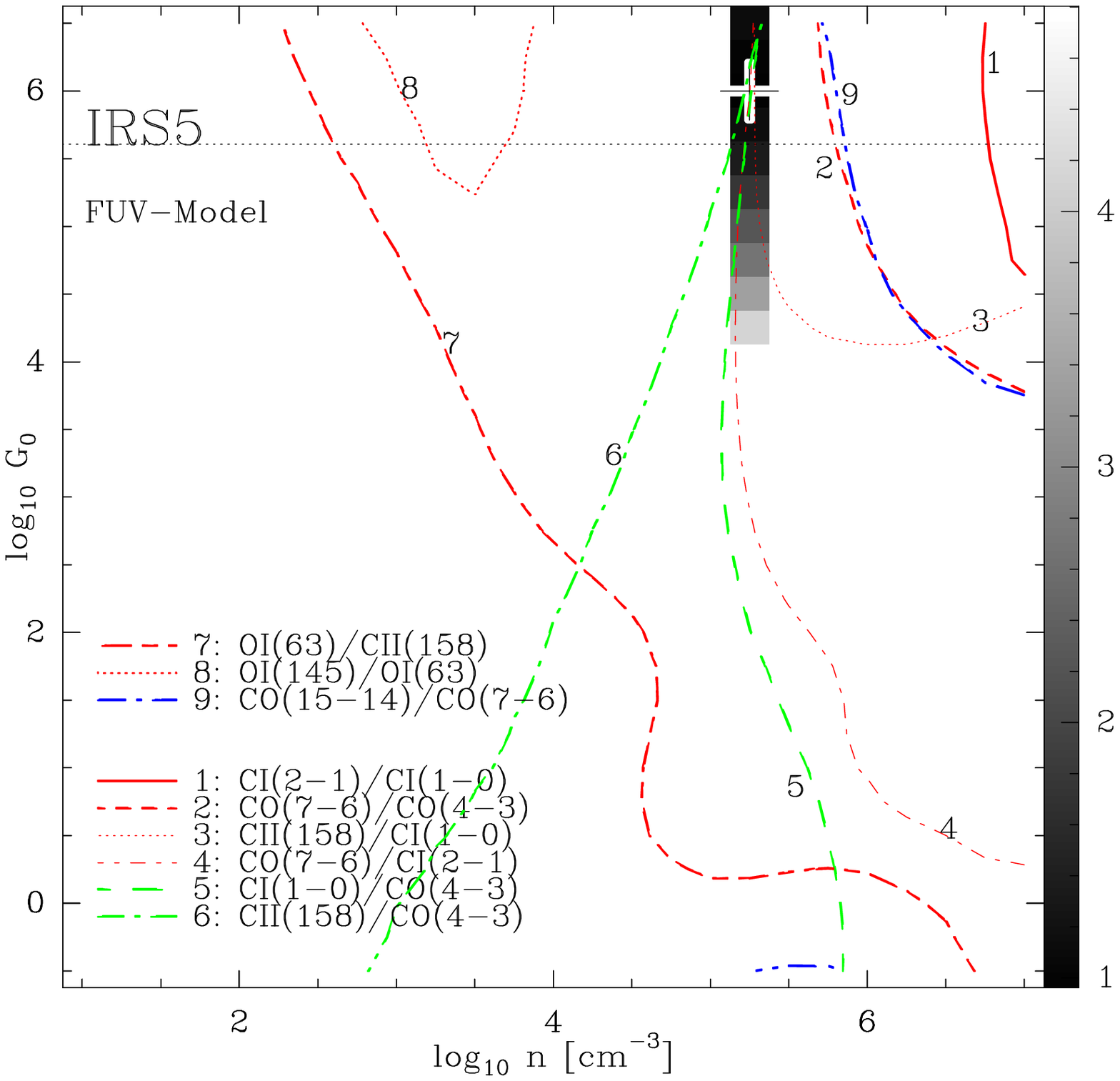}\nobreak
\includegraphics[angle=0,totalheight=6cm]{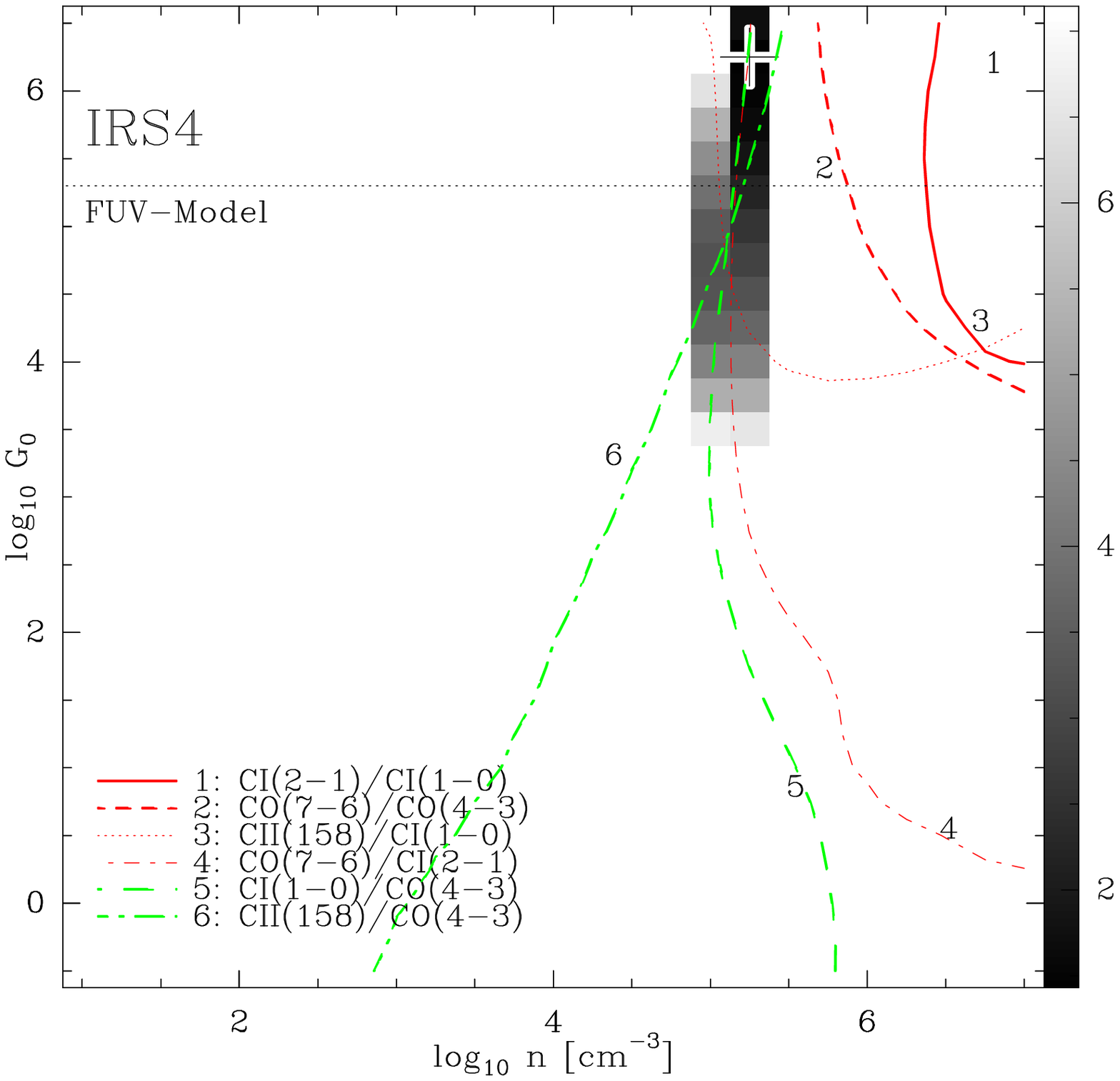}\break
\includegraphics[angle=0,totalheight=6cm]{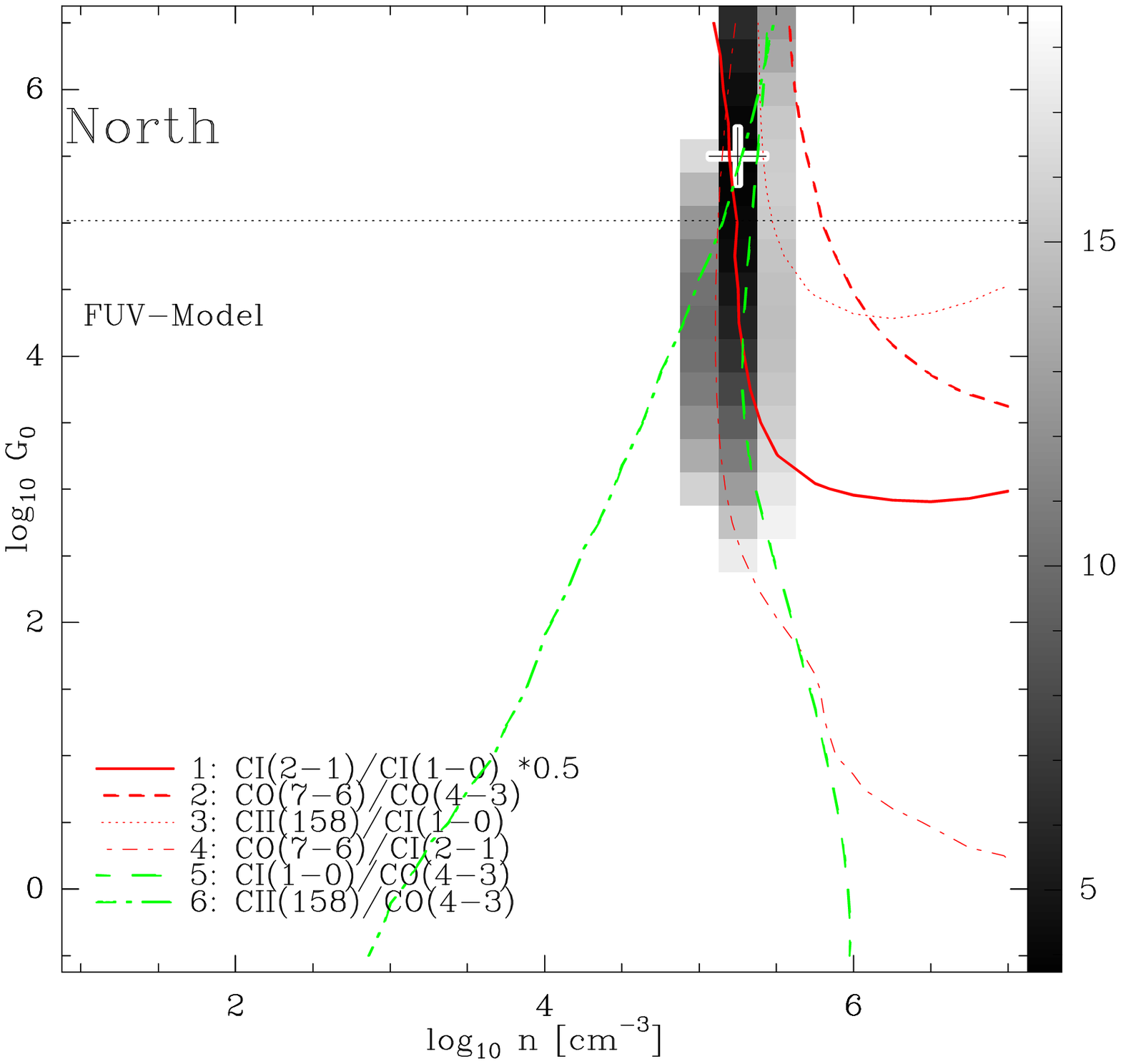}\nobreak
\includegraphics[angle=0,totalheight=6cm]{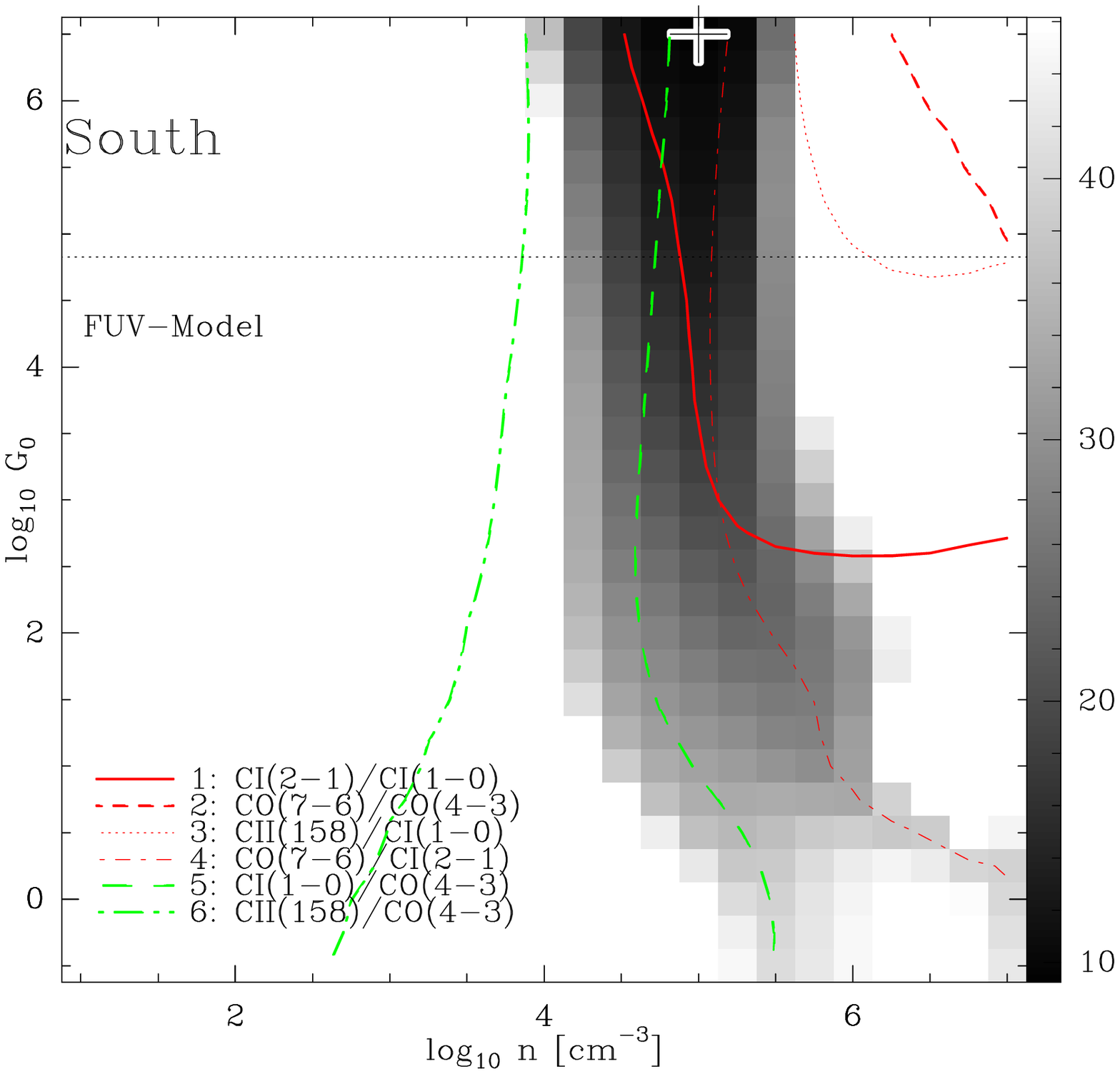}\break
%
\caption{
  Results from the PDR model of \citet{kaufman1999} at the four
  selected positions in W3\,Main. Contours show the observed ratios of
  line integrated intensities as a function of FUV flux in units of
  the Habing field $G_0$ and the H$_2$ volume density $n$. We use the
  \CII\ data from KAO \citep{howe1991}. The straight horizontal lines
  show the FUV field modelled from the embedded OB stars and smoothed
  to the resolution of the line data, i.e. $75''$
  (Table\,\ref{tab-carbon-content}). Greyscales denote the reduced
  $\chi^2$ derived from the observed line ratios, weighted with the
  uncertainty of 20\%, and not taking into account the ratios of
  \OI(63$\,\mu$m) and CO 15--14 at IRS\,5. The cross marks the best
  solution $\chi_{\rm min}^2$.  Greyscales vary between 1 and 5 times
  $\chi_{\rm min}^2$.
\label{fig-pdr-kaufman}}
\end{figure*}

\begin{figure}[htb]
\centering
\includegraphics[angle=0,totalheight=6cm]{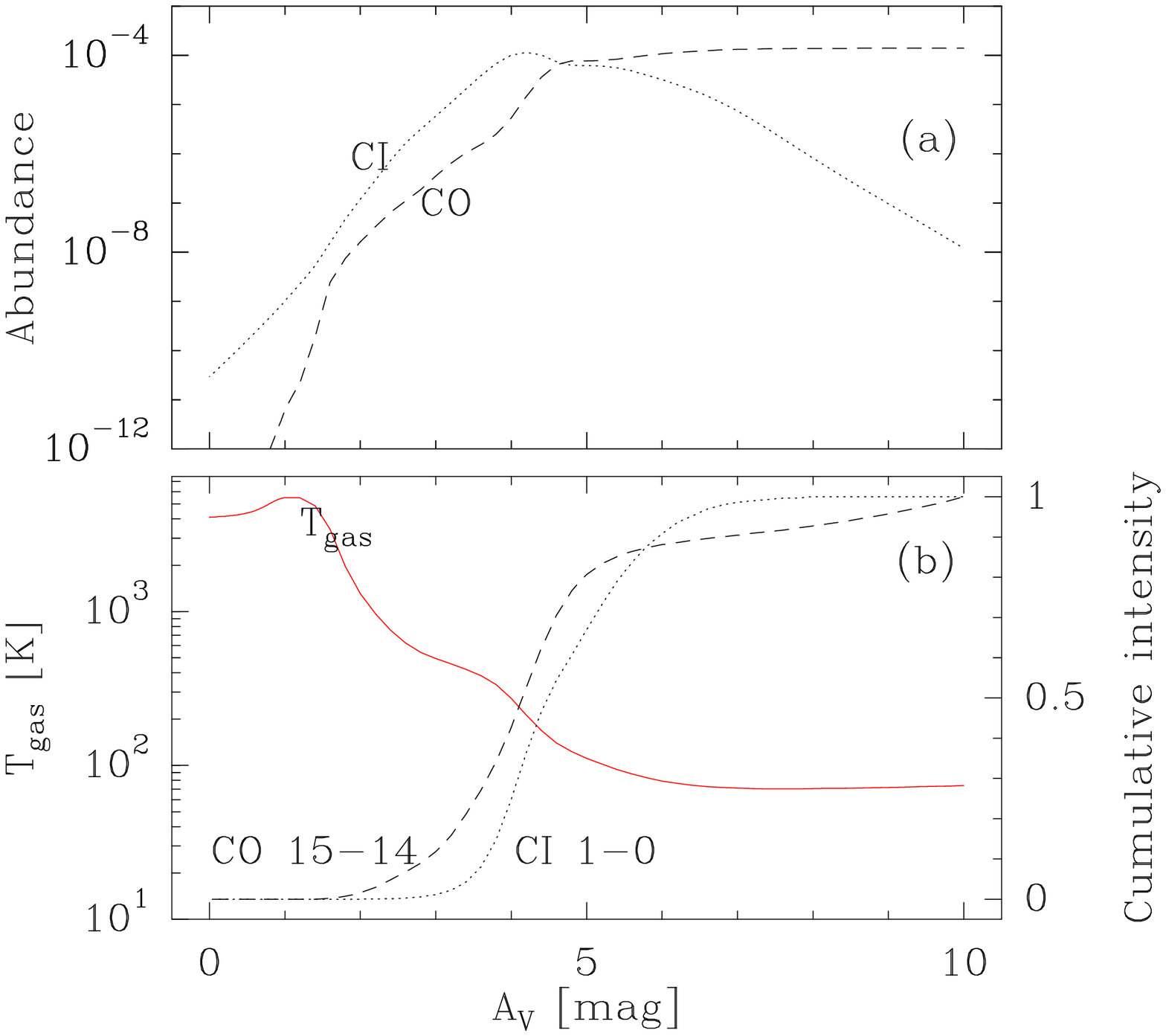}
\caption{  
\label{fig-pdr-kaufman-av-co-ci}
Variation of physical parameters with $A_V$ for a PDR model
\citep[][Wolfire priv. comm.]{kaufman1999}with
$n=1.8\,10^5$\,cm$^{-3}$ and $G_0=10^6$ which is characteristic for
IRS\,5 (cf.\,Table\,\ref{tab-pdr-results}).
%
%
{\bf (a)} Local abundances $n(X)/(n(H)+2n(H_2))$ of atomic carbon and
CO.  The CI abundance peaks at $A_V\sim4.2\,$mag reaching
$1.2\,10^{-4}$ while the CO abundance rises steadily with $A_V$
reaching an abundance of $1.4\,10^{-4}$ at $A_V=10\,$mag.
{\bf (b)} The gas temperature (drawn line) varies between 4000\,K at
the surface and $\sim70$\,K at $A_V=10$\,mag.  The contours labeled CO
15-14 and CI 1-0 show the line intensity integrated along $A_V$ from
the outside in to greater depths and normalized to the integrated
intensity at $A_V=10\,$mag. The CO 15-14 line stems from regions
slightly further out and slightly warmer than the region where CI 1-0
is emitted.  The extinction upto which half of the intensity is
emitted, is 4.2\,mag for CO 15-14 and 4.6\,mag for CI 1-0.  At these
extinctions, the gas temperatures are 213\,K and 140\,K, respectively.
}
\end{figure}

\begin{table*}
\caption[]{Results of detailed PDR modelling at the four selected positions in W3\,Main. 
%
The minimum $\chi^2$ was used to derive an  estimate of the best fitting clump density $n_{\rm cl}$
and UV Flux. $T_{\rm surf}$ is the surface temperature of the corresponding clump
predicted by the PDR model.
The average density $n_{\rm av}$, the volume filling factor $\phi_V$, 
the average clump diameter $D_{\rm cl}$, and the number of clumps
 are derived using the PDR model results and the total beam averaged 
H$_2$ column density resulting from 
the above LTE analysis (Table\,\ref{tab-carbon-content}). 
The area filling factor $\Phi_A$ is the ratio of observed \CII\ intensities over
the modelled intensities.
}
\label{tab-pdr-results}
\begin{tabular}{cllllllllll}
\noalign{\smallskip} \hline \hline \noalign{\smallskip}
  $\Delta\alpha/\Delta\delta$ & $\chi_{\rm min}^2$ & $n_{\rm cl}$     & UV Flux & $T_{\rm surf}$ 
  & $n_{\rm av}$ & $\phi_V$ & $D_{\rm cl}$ & $N_{cl}$ & $\phi_A$ \\ 
         $['','']$         &          & [$\log$(\cmcub)] & [$\log(G_0)$] & [K] 
  & [\cmcub]     &          & [pc]         &           & \\
\noalign{\smallskip} \hline \noalign{\smallskip}
 $0,0$ (IRS\,5)        & 1.0 & 5.25 & 6    & 3970 & $1.5\,10^4$ & 0.09 & 0.05 & 404 & 1.5 \\
 $-55.0,27.5$ (IRS\,4) & 1.4 & 5.25 & 6.25 & 4070 & $1.6\,10^4$ & 0.09 & 0.07 & 188 & 1.2 \\
 $0,82.5$ (North)      & 3.7 & 5.25 & 5.5  & 3640 & $5.8\,10^3$ & 0.03 & 0.03 & 467 & 0.8 \\
 $0,-82.5$ (South)     & 9.2 & 5.00 & 6.5  & 3480 & $7.3\,10^3$ & 0.07 & 0.05 & 340 & 1.2\\
\noalign{\smallskip} \hline \noalign{\smallskip}
\end{tabular}
\end{table*}

\subsection{Detailed PDR modelling of the submillimeter and far-infrared emission}
\label{chap-pdr-model}

We have used the results of PDR models of \citet{kaufman1999} to
compare with the observed line ratios. These models assume a constant
density of a 
 semi-infinite
plane-parallel slab illuminated by the interstellar radiation field
and improve on the older models of \citet{tielens1985}. PAHs are found
to dominate the grain-photoelectric heating and are therefore
included.  $^{13}$C is not included in the chemical network.  The
model computes a simultaneous solution for the chemistry, radiative
transfer, and the thermal balance in PDRs.
%

\subsubsection{Estimating density and FUV field from the line ratios}

We have compared the models with selected line flux ratios of the PDR
tracers: CO (4--3, 7--6), \CI\ (1--0, 2--1), \CII\ 158$\mu$m, and, at
IRS\,5, \OI (63$\mu$m, 145$\mu$m) and CO (15--14).
Figure\,\ref{fig-pdr-kaufman} shows the detailed comparison at the
four selected positions in W3\,Main (cf.  Fig.\,\ref{fig-irs4and5},
Tables\,\ref{tab-wco-positions}, \ref{tab-wco-ratios},
\ref{tab-carbon-content}, \ref{tab-isodata}).  The plots span a
density range from $n=10$ to $10^7$\,\cmcub\ and a FUV flux range from
$G_0=0.3$ to $3\,10^6$,
i.e. the full range of the Kaufman models.
Horizontal dashed lines mark the beam averaged FUV flux estimated
above in Sect.\,\ref{chap-fuv-field}
 from the spectral types of the OB stars.

We do not expect to find one solution of $G_0, n$ for each observed
position, i.e. an intersection of contours at one point. This is
because, first, the observational calibration error of 20\% of the
ratios and, second, the line emission surely cannot be entirely
described by slabs of only one density.
%
%
The analysis given here will however allow to derive average clump
densities, average impinging FUV fields, and several other average
clump parameters. Indeed, the line ratio contours shown in
Figure\,\ref{fig-pdr-kaufman} for the four representative positions do
not all intersect. However, the observed four ratios of \CII/\CI\ 
1--0, CO 7--6/\CI\ 2--1, \CI\ 1--0/CO 4--3, and \CII/CO 4--3 do almost
intersect at $n_{\rm cl}=2\,10^5$\,\cmcub\ and $G_0>10^5$ at the three
positions IRS\,5, IRS\,4, and North. 
 The ratios of CO over \CI\ lines largely determine the densities
  and are rather independent of $G_0$. The high FUV fields are largely
  determined by the \CII\ over \CI\ ratios and also by the \CII\ over
  CO 4--3 ratios. 
%
We find still higher densities of upto $10^7$\,\cmcub\ from the two
ratios of \CI\ 2--1/1--0 and CO 7--6/4--3 at all positions.  

At IRS\,5, we also obtained the CO 15--14/7--6 ratio. Very similar to
the CO 7--6/4--3 ratio, it indicates high densities of
$\sim6\,10^5$\,\cmcub\ for $G_0>10^5$.  However, these ratios are
still consistent with the best fitting solutions within a deviation of
better than 50\%.
%
 Figure\,\ref{fig-pdr-kaufman-av-co-ci} shows the variation of
  abundances, temperatures, and intensities with optical extinction
  for the best fitting PDR model at IRS\,5: $n=1.8\,10^5$\,cm$^{-3}$,
  $G_0=10^6$, i.e. $G_0/n=5.5$\,cm$^3$
  (cf.\,Table\,\ref{tab-pdr-results}). It shows that the bulk of CO
  15--14 emission stems from a region of $\sim210\,$K temperature
  while the bulk of CI 1--0 emission stems from a region of
  $\sim140$\,K.  While the detailed PDR modelling shows a strong
  temperature gradient, the characteristic CO temperature of 210\,K is
  only slightly higher than the results of the single component
  escape-probability modelling of the mid and high-$J$ CO lines
  described in sec.\,\ref{sec-esc-prob}
  (Fig.\,\ref{fig-co-cooling-irs5}).  There, we had fitted
  temperatures and densities of between 140\,K at $10^5$\,cm$^{-3}$
  and 200\,K at $3\,10^4$\,cm$^{-3}$.  However, a homogeneous model
  with 200\,K at $n=10^5$\,cm$^{-3}$ would be predicting stronger CO
  lines than considered in Section\,\ref{sec-esc-prob}. Thus, with
  regard to CO emission, the esc. prob. model and the PDR model do not
  agree with each other completely.
  
  The LTE analysis of the CI line ratio in
  section\,\ref{chapter-ci-line-ratios} indicated CI excitation
  temperatures of $>120\,$K at IRS\,5, in rough agreement with the
  detailled PDR modelling.



\subsubsection{The \OI($63\,\mu$m) line is optically thick}
\label{sec-oi63}

From ISO/LWS at IRS\,5, we also obtained ratios of
\OI(63\,$\mu$m)/\CII\ and \OI(145\,$\mu$m)/(63\,$\mu$m).  Assuming a
FUV field of at least $G_0=10^5$, these ratios indicate low densities
of less than $10^4$\,\cmcub, clearly deviating from the above
scenario.  However, the \OI($63\,\mu$m) line is optically thick all
over the parameter space \citep{kaufman1999} and thus probably
strongly affected by self-absorption of the cold foreground gas
component also visible in the low-$J$ and mid-$J$ CO lines. 
 The line strengths of the CO 4--3 and 3--2 lines are reduced by a
  factor of 1.5 because of self-absorption
  (Fig.\,\ref{fig-irs5-gauss}). Decreasing the
  \OI(145\,$\mu$m)/(63\,$\mu$m) ratio by a factor of 2 would lead to a
  contour intersecting with the best fitting $G_0, n$ solution of the
  other ratios. However, the \OI(63\,$\mu$m)/\CII\ would need to be
  increased by a factor of 15. Thus, the \OI(63\,$\mu$m) line is not
  consistently described in the current model.
%
Velocity resolved observations to study the structure of FIR line
shapes and thus confirm self-absorbed profiles will become possible
with GREAT/SOFIA \citep{guesten2003} in the near future.

\subsubsection{The quality of the fits}

To quantify the correspondance of the ratios, we have plotted in
Figure\,\ref{fig-pdr-kaufman} the $\chi^2$ per degree of freedom, i.e.
the reduced $\chi^2$, of the observed ratios in greyscales. To allow
for a comparison between the four positions, we did not take into
account here the ratios of \OI\ and CO 15--14 which were only obtained
at IRS\,5.  For the six remaining ratios, the degree of freedom is
$5-2$.
 \footnote{Only 5 ratios are independent: the ratio CO 7-6/CI 2-1 can be 
derived from CI 2-1/CI 1-0, CI 1-0/CO 4-3, CO 4-3/CO 7-6.}
The best fitting solution $\chi_{\rm min}^2$ is marked by a cross and
listed in Table\,\ref{tab-pdr-results}. A value of $\chi_{\rm min}^2$
near 1 is found at IRS\,5 and IRS\,4 which signifies that the ratios
agree well within their observational errors. The agreement is worse
at the two other positions.

The best fitting FUV fluxes of the PDR models are surprisingly high,
i.e.  upto $G_0=2\,10^6$, though with a higher uncertainty compared to
the best fitting densities. At all positions, the fitted FUV flux is
higher than the beam averaged, purely geometrically attenuated FUV
fluxes estimated above from the luminosity of the embedded OB stars.

%

\subsubsection{Line and FIR cooling at IRS\,5 revisited}
\label{sec-total-fluxes}

 The line fluxes observed with KOSMA and ISO/LWS at IRS\,5 had
  already been discussed in Chapter\,\ref{sec-line-cooling}. Here,
  these fluxes are compared with the absolute fluxes predicted by the
  best fitting PDR model at IRS\,5 (cf. Table\,\ref{tab-pdr-results}).
  Table\,\ref{tab-total-fluxes} shows that model and observations are
  in agreement to within 50\% for the total fluxes of CO, \CII, and
  \CI.  The \CII\ flux observed with the KAO listed here is a factor
  of 1.5 larger than the flux observed with ISO which equals the PDR
  model result.
  
  As discussed in Sec.\,\ref{sec-oi63}, the \OI\ $63\,\mu$m line is
  predicted to be much stronger than observed.  The total observed
  \OI\ cooling is only 10\% of the modelled flux.
  
  As already noted above, the FUV flux derived by fitting the line
  ratios to the PDR model is $G_0=10^6$, surprisingly high relative to
  the beam averaged FUV flux derived from the OB stars, $G_0=4\,10^5$,
  and higher than the FUV flux derived from the beam averaged FIR
  field, $G_0=5\,10^4$ (Sec.\,\ref{sec-grey-body-fit}).
    Correspondingly, the FIR flux derived from the modelled $G_0$ via
    $I_{\rm FIR} = 2\times1.6\,10^{-3} (4\pi)^{-1} G_0$\,erg
    s$^{-1}$cm$^{-2}$sr$^{-1}$ \citep{kaufman1999}, is high, $I_{\rm
      FIR}=260$\, erg s$^{-1}$ cm$^{-2}$ sr$^{-1}$, i.e. a factor $20$
    higher than the observed flux. 

  This indicates that the line emission is mainly excited in the dense inner
  regions near the embedded OB stars where the FUV flux is higher. The FIR
  continuum on the other hand also stems from the colder outer surroundings
  thus leading to a lower average FIR intensity.

\begin{table}
\begin{center}
\caption{\label{tab-total-fluxes}
  Observed fluxes at IRS\,5 in units of $10^{-3}$ erg s$^{-1}$
    cm$^{-2}$ sr$^{-1}$ compared to the best fitting PDR model with
    $n=1.8\,10^5$\,cm$^{-3}$, $G_0=10^6$. The total gas cooling is
    $\Lambda_{\rm gas}=\Lambda_{\rm CII}+\Lambda_{\rm CI}+\Lambda_{\rm
      CO}+\Lambda_{\rm OI}$. $^a$ The total ``observed'' CO cooling
    $\Lambda_{\rm CO}$ was derived from the esc.  prob. model fitted
    to the observed fluxes (cf.  Chapter\,\ref{sec-line-cooling}). }
\begin{tabular}[h]{lrrrrrr}
\hline \hline
          & $\Lambda_{\rm CII}$ 
                & $\Lambda_{\rm CI}$ 
                     & $\Lambda_{\rm CO}$ 
                          & $\Lambda_{\rm OI}$ 
                               & $\Lambda_{\rm gas}$ 
                                     & $\Lambda_{\rm FIR}$ \\
\hline 
Observed  & 2.2 & 0.04 & 4.9$^a$ &  9.1 &  15.5 &  13500 \\
PDR-Model & 1.5 & 0.03 & 7.3     & 93.7 & 103.0 & 260000 \\
\hline 
Ratio     & 1.5 & 1.3  & 0.7     &  0.1 & 0.15  & 0.05 \\
\hline
\end{tabular}
\end{center}
\end{table} 

\subsubsection{A clumpy PDR model}

Using these results in combination with the total beam averaged H$_2$
column density resulting from the LTE analysis at the four positions
(Table\,\ref{tab-carbon-content}), allows to derive more physical
parameters characterizing the line emitting gas. We assume here, that
line emission arises in individual clumps immersed in an interclump
medium which is not emitting in the observed transitions.  
 Motivated by the morphology of the KOSMA maps shown in
  Figure\,\ref{midj}, we assume conservatively that the line of sight
  extent of the cloud is $D_{\rm cloud}=0.84$\,pc ($75''$). This
  allows to use the total column density to derive a rough estimate of
  the average volume density $n_{av}$ per beam.  
The ratio of $n_{av}$ with the average clump density $n_{cl}$ derived
above, gives the volume filling factor of the clumps $\Phi_V$.  The
area filling factor $\Phi_A$ of the clumps can be estimated from the
ratio of observed intensities $I_{\rm obs}$ over the modelled
intensities $I_{\rm mod}$ - it can be larger than unity for optically
thin emission. The clump diameter $D_{\rm cl}$ can be estimated from
$D_{\rm cloud}\,\Phi_V/\Phi_A$.  The number of clumps is then given by
$\Phi_V (D_{\rm cloud}/D_{\rm cl})^3$.


Table\,\ref{tab-pdr-results} summarizes the resulting quantities at
the four positions. Results are much less certain for the southern
position than for IRS\,5.  Only less than 10\% of the volume is filled
with a few hundred clumps per beam area. Clumps have a typical
diameter of 0.05\,pc, i.e. $4.4''$, and are thus unresolved. The
resulting typical clump mass is $M=n_{\rm cl}\,4/3\,\pi\,(D_{\rm
  cl}/2)^3=0.44\,$\msol.

The area filling factor derived from \CII\ is $\sim1\pm0.5$. Area
filling factors and clump diameters are not significantly changed when
e.g.  using the modelled \CI\ 1--0 intensities instead. Another reason
for discrepancies between the modelled and the observed intensities
are geometrical effects. \citet{kaufman1999} assume a face-on
geometry, a viewing angle of $<90^\circ$ would lead to an increase in
the intensity of optically
 thin lines.

 We cannot exclude the possibility that a fraction of the CI and
  CII emission stems from the interclump medium since the CI and CII
  transitions have critical densities of only $5\,10^2$ to
  $3\,10^3$\,cm$^{-3}$ \citep[cf.][]{kaufman1999}.

These results confirm the results of \citet{howe1991} which had been
based solely on the \CII\ map and on the estimated FUV field. They had
found that a cloud clumpy model can be adjusted to be roughly
consistent with the observed \CII\ morphology.  Their best fitting
model is very similar to ours: their clumps have a density of $n_{\rm
  cl}\sim10^5\,$\cmcub, the volume filling factor is
$\Phi_V\sim0.05-0.07$.


 Further indirect support for our clumpy source model comes from
  the analysis of CS 2--1, 5--4, and 7--6 FCRAO and KOSMA small maps 
  of W3\,Main by \citet{ossenkopf2001}. The observed line profiles are best
  fitted by an average density of $0.7\,10^4$\,cm$^{-3}$ within a
  radius of 1\,pc while the minimum central density reaches
  $5\,10^6$\,cm$^{-3}$, higher than the clump density we have
  derived from the PDR analysis, but similar to the high densities
  inferred from the \CI\ 2--1/1--0 and CO 7--6/4--3 ratios. 
  
  Direct evidence for the existence of very dense and small molecular
  clumps is delivered by VLA maps of Ammonia in W3\,Main at $3''$
  resolution by \citet{tieftrunk1998b}. These show several clumps of
  $\sim0.02$\,pc radius and densities of a few $10^7$\,cm$^{-3}$ in
  the vicinity of IRS\,5 and IRS\,4.  
  
  Still higher densities at temperatures of almost 600\,K are known to
  exist on scales of less than $0.01$\,pc ($<2000$\,AU, $<1''$) in the
  immediate vicinity of IRS\,5 from the absorption study of
  \citet{mitchell1990}.

Our scenario of many small, high density clumps with a small volume,
but a high area filling factor should nevertheless not be taken too
literal. It simply implies that the cloud structure is very far off a
homogeneous cloud, but shows a very much broken up structure with high
density contrasts and steep density gradients within the cloud.

\section{Summary} 

\begin{enumerate}
  
\item We have mapped the W3\,Main region in the two fine structure
  lines of atomic carbon and in several mid-$J$ CO transitions using
  KOSMA.  All maps show a peak of integrated intensities near IRS\,5
  and IRS\,4 with a steady drop of intensities in all directions.
  Spectral line maps show structure down to the spatial resolution
  limit of $>42''$, i.e. $>0.5\,$pc. To obtain a common spatial
  resolution, all data discussed below are at a common resolution of
  $75''$ (0.84\,pc). In addition, we have obtained ISO/LWS data at
  IRS\,5.
  
\item We find a constant CO 7--6/4--3 line integrated intensity ratio
  of $\sim1$ at all positions in W3\,Main after correcting for
  self-absorption seen in the $^{12}$CO lines in the vicinity IRS\,5.
  Assuming thermalized emission, this ratio indicates excitation
  temperatures in excess of 85\,K and densities in excess of
  $\sim10^6\,$\cmcub\ for the gas emitting these mid-$J$ CO lines.
  A more detailed non-LTE analysis however reveals that densities
  of $\sim10^5$\,\cmcub\ are sufficient to explain the observed 
  CO cooling fluxes.

\item The observed high \CI\ 2--1/1--0 line ratios of between 1.3 and
  2.9 indicate optically thin carbon emission and excitation
  temperatures of about 100\,K. 

  
\item LTE analysis of carbon column densities show that the \CIw/CO
  abundance ratios stays constant at $\sim0.11$. The fractional
  abundances of \CIIw/\CIw/CO are $\sim20:10:70$ at IRS\,5, rising to
  $\sim32:7:61$ in the south of W3\,Main.
  
\item A model of the FUV field distribution, derived from the embedded
  OB stars, assuming pure geometrical attenuation, ignoring the cloud
  density structure, and smoothed to $75''$ resolution, shows a peak
  field at IRS\,5 of $G_0=10^{5.5}$ dropping to values below
  $10^{3.5}$ in the outskirts of W3\,Main.
%

\item The north-south orienation of the \CII\ emission
  \citep{howe1991} is not seen in the maps of \CI\ and CO emission.
  Nor is it seen in the modelled FUV field distribution.
  
\item At IRS\,5, the (average) dust temperature of the FIR emission
  detected by ISO/LWS is 53\,K with a spectral index of $\beta=1.2$.
  The \OI\ 63\,$\mu$m line is a factor 4 stronger than the \CII\ 
  158$\,\mu$m line. The \CII/FIR flux ratio is 
   0.01\%.
  We estimate that less than 1\% of the \CII\ emission stems from the
  \HII\ regions in W3\,Main. The relative contributions to the total
  gas cooling are $\Lambda_{\rm CO}=31.6\%$, $\Lambda_{\rm O}=58.7\%$,
  $\Lambda_{\rm CII}=9.7\%$, $\Lambda_{\rm C}=0.3\%$.  $\Lambda_{\rm
    CO}$ refers to the total CO cooling modelled with the escape
  probability radiative transfer model in
  Sec.\,\ref{sec-line-cooling}. All other numbers refer to the
  observed values.  The ratio of $\Lambda_{\rm gas}/\Lambda_{\rm
    dust}$ is
    0.1\%.
  
\item We have used the PDR model of \citet{kaufman1999} to estimate
  the average densities of the emitting region and the impinging FUV
  fields from the observed line ratios at four selected positions. The
  \OI(63$\mu$m) line is probably optically thick and strongly affected
  by self-absorption in the colder foreground material at IRS\,5, and
  thus discarded from the further analysis. The remaining six line
  ratios indicate a common solution of $G_0\sim10^6$ and $n_{\rm
    cl}\sim2\,10^5$\,\cmcub\ at all four positions. The reduced
  $\chi^2$ is low at IRS\,5, IRS\,4, and the northern position, but
  much higher at the Southern position where no well defined common
  solution is found. We have assumed that all line emission arises in
  individual clumps immersed in an interclump medium which is not
  emitting in the observed transitions. We find that the emitting gas
  consists of a few hundred clumps per 0.84\,pc beam, filling between
  3 and 9\% of the volume, with a typical clump radius of 0.025\,pc
  ($2.2''$), and typical mass of 0.44\,\msol.
  
\item The line ratios of \CI\ 2--1/1--0 and CO 7--6/4--3 are
  consistent with very high clump densities of upto $10^7$\,\cmcub\ in
  the framework of the PDR model. This finding may be explained by a
  more detailed model of clumps having a variation of densities (and
  masses). A more detailed model of W3\,Main would also need to take
  into account absorption by foreground gas. 
  

\end{enumerate}

Although the discussed model scenario is not able to reproduce each
detail of the line intensities and their variation within the cloud,
it is 
remarkable
%
that the scenario of a UV penetrated clumpy cloud is able to reproduce
the relative (and to a large degree also the absolute) intensities of
some dozen lines simultaneously with only a relatively small set of
model parameters such as the clump average density and size, most
other parameters being constrained by the UV intensity distribution
and the temperature and chemical gradients calculated through the PDR
models.

\begin{acknowledgements}
  We thank the anonymous referee for thoughtful comments which helped
  to improve this paper.  The CO $3-2$ data set was observed by Anja
  M\"uller as part of her diploma thesis at the Universit\"at zu
  K\"oln. 
  We thank Mark Wolfire for providing additional information
  concerning the PDR model we have used.
  The KOSMA 3m submillimeter telescope at the Gornergrat-S\"ud is
  operated by the University of Cologne in collaboration with the
  Radio-Astronomical Institut at Bonn University, and supported by
  special funding from the Land NRW.  The observatory is administrated
  by the International Foundation Gornergrat \& Jungfraujoch.
\end{acknowledgements}

\bibliographystyle{aa}
\bibliography{aamnem99,p_w3_bib}


\end{document}